\documentclass[a4paper, amsfonts, amssymb, amsmath, reprint, nofootinbib, twoside, notitlepage]{revtex4-1}

\usepackage[centering,hmargin=2cm,vmargin=2cm,lmargin=1.7cm,rmargin=1.7cm]{geometry}
\usepackage[T1]{fontenc}
\usepackage{amsmath,amstext,amssymb,mathtools}
\usepackage{bm}
\usepackage{graphicx}
\usepackage{subcaption}
\usepackage[hidelinks]{hyperref}
\usepackage{cleveref}
\usepackage{enumitem}
\usepackage{float}
\usepackage{booktabs}

\begin{document}

\title{Fluxtube Bouquets and Type-1.5 Clustering in Superfluid Neutron Star Cores}

\author{\href{https://orcid.org/0009-0003-4812-1047}{Adarsh Karekkat}}
\author{\href{https://orcid.org/0009-0004-7538-6064}{Gabriele Montefusco}}
\author{\href{https://orcid.org/0000-0002-5470-4308}{Marco Antonelli}}
\email[Email: ]{antonelli@lpccaen.in2p3.fr}
\affiliation{CNRS/IN2P3, Laboratoire de Physique Corpusculaire de Caen, 14050 Caen, France}

\begin{abstract}
	We study mesoscopic configurations of a neutron superfluid coupled to a proton superconductor in the outer core of a neutron star. The condensates are described by a two-component Ginzburg-Landau free energy with local couplings, neglecting genuine phase-gradient entrainment. In two spatial dimensions, we minimize the free energy using quasi-periodic boundary conditions and constrained phase-imprinting calculations to study vortex-fluxtube and fluxtube-fluxtube interactions.
	We find that, for locally attractive couplings in the free energy, vortex-fluxtube overlap is energetically favoured and several pre-existing proton fluxtubes can bind around a neutron vortex, forming finite vortex-centred aggregates that we call fluxtube bouquets. These bouquet configurations may become so dense that a vortex can effectively accommodate several quanta of magnetic flux.
	We also confirm the possible presence of a type-1.5-like regime and find that it survives in the zero-entrainment regime considered here. In this type-1.5 regime, the fluxtube-fluxtube interaction is repulsive at short distances and attractive at intermediate distances, leading to self-assembled clusters while the individual fluxtubes remain topologically distinct. 
	Possible implications for dissipative coupling and transport in neutron stars are discussed.
\end{abstract}

\maketitle

\section{Introduction}
\label{sec_introduction}

Superfluid and superconducting phases are expected to play an important role in several aspects of neutron star evolution 
\citep{haskell_super,chamel_review2017JApA,graber_NSlab_2017IJMPD}.
In particular, neutron superfluidity is central to the current interpretation of pulsar glitches \citep{AMP_arxiv_2023,Antonopoulou_review_2022,Zhou_2022Univ} and may also contribute to pulsar timing noise \citep{Meyers2021MNRAS,antonelli2023MNRAS,antonelli2025PASA}, while both neutron superfluidity and proton superconductivity are invoked to interpret and model thermal evolution \citep{potekhin_2015SSRv,ginzales2015MNRAS,Das_burgio_2024PhRvD}, magnetic field evolution \citep{Ruderman1998,alpar2017JApA,gusakov2017PhRvD,pons2026LRCA} and can also affect thermo-chemical evolution through rotochemical heating~\citep{ginzales2015MNRAS,rodriguez_reisenegger_2026}.

In the outer core, the neutron component is expected to form a superfluid threaded by quantized vortices, while the proton component can form a superconductor in which magnetic flux is carried by quantized fluxtubes, at least in regions where the superconducting state is type-II-like \citep{Baym1969_typeII,sedrakian2019EPJA}. The interaction between these two families of topological defects can therefore couple the rotational dynamics of the neutron superfluid to the charged component and magnetic field, making vortex-fluxtube interactions relevant to several of the phenomena discussed above.

The astrophysical importance of vortex-fluxtube interactions was recognized in early work on recycled millisecond pulsars. Soon after their discovery, accretion-driven recycling was proposed as their formation mechanism \citep{Srinivasan1,Alpar1982}, but their comparatively weak magnetic fields subsequently motivated the idea that neutron vortices can interact with proton fluxtubes during spin evolution and advect magnetic flux outward \citep{Srinivasan2}. 
More generally, vortex-fluxtube pinning and its macroscopic consequences have been considered extensively in neutron star core models \citep{Sedrakyan1991JETP,Srinivasan2,Ruderman1998,alpar2017JApA,sourie2020_pinningCore,souriechamel2020}. 

Long-period precession has also been used to argue that a strongly pinned type-II superconducting core may be difficult to reconcile with observations \citep{Link2003,Link2006,Stairs2000}, although imperfect pinning and other mechanisms have been explored \citep{Sedrakian_precession_1999,Kitiashvili2008,Goglichidze2018,Charbonneau2007}. Conversely, it was proposed that, if the crustal angular-momentum reservoir is insufficient for some large glitches, the core may contribute through vortex-fluxtube coupling \citep{andersson+2012,chamel2013}. Whether the crust alone is sufficient, and how strongly entrainment reduces the effective mobile superfluid fraction, remains debated, and recent results (see \citep{urban_sf_fraction,Chamel2025}) show that, considering the current uncertainty on the neutron star equation of state and crust modelling, the angular momentum reservoir in the inner crust is typically enough to explain glitches~\citep{burrello2025PhRvC,klausner2026PhRvC}. This, however, does not fully rule out the implications of vortex-fluxtube pinning in glitches, as it is a conclusion based solely on the observed glitch activity averaged over long observational baselines; see, e.g., the discussion in \citep{AMP_arxiv_2023,montoli_universe}. On the other hand, the Bayesian analysis of the 2016 glitch in the Vela pulsar may carry a clue that some involvement of the outer core is necessary, regardless of the strength of crustal entrainment~\citep{montoli2020AA}.

At the microscopic and mesoscopic levels, superfluid-superconducting phenomenology in neutron star matter has often been discussed using variations of the two-condensate phenomenological Ginzburg-Landau model introduced by \citet{alpar1984ApJ}. Such models have been extensively used in the literature. Of particular relevance to the present study are \citep{alford_good_PRB,haber2017prd,Wood_2022Univ} and \citep{Drummond2017I,DrummondL2018II,melatos_III_2023}; see also~\citep{mendell1991,Sedrakyan1991JETP,Sedrakian1995ApJ, sedrakian1997MNRAS,Babaev:2002wa,Buckley-etal04-a,Buckley-etal04-b,Jones06,Charbonneau2007,SinhaSedrakian2015,KobyakovSupercond,Shukla_PRD_2024,hattori2026arXiv,hattori2025}.
In particular, \citet{alford_good_PRB} showed that a coupled superfluid can shift the type-I/type-II boundary and, in some regions of parameter space, can favour multi-quantum fluxtubes. 
Using a similar two-component phenomenological Ginzburg-Landau model, \citet{haber2017prd} found that the transition region can also support clustered type-1.5 fluxtube phases, analogous to the  type-1.5 behaviour that was already known in multicomponent terrestrial superconductors \citep{Babaev2005prb,Moshchalkov2009,Reimann2015}. Following a similar line, \citet{Wood_2022Univ} subsequently argued, including current-current entrainment couplings, that type-1.5 superconductivity may occur over a substantial part of the outer core.

A useful tool for studying Ginzburg-Landau topological defects in a finite computational domain is provided by quasi-periodic boundary conditions (QPBC). A first numerical implementation for a single superconductor was given in two dimensions by \citet{doria_1988}. For neutral superfluids, analogous boundary conditions were later used to simulate two-dimensional vortex lattices by \citet{Mingarelli_2016,Mingarelli2018}. \citet{Wood2019PhRvB} formulated the construction systematically and extended it to three dimensions, showing that the same framework applies both to vortices in a neutral Gross-Pitaevskii condensate and to fluxtubes in a charged Ginzburg-Landau condensate. QPBC have since been used in single neutral condensates to study vortex depinning and bulk vortex dynamics \citep{Liu2024,Magistrelli_2026pra}. They were also applied by \citet{Wood_2022Univ} to a coupled neutron-proton Ginzburg-Landau model to study the phase diagram and the arrangements of fluxtubes in a vortex-free neutron condensate. Here we complement this study by using QPBC with topological defects simultaneously present in both condensates, allowing neutron vortices and proton fluxtubes to coexist and relax within the same computational cell and enabling us to extract pinning energies.

The goal of this paper is thus to study local configurations and the mesoscopic interaction between neutron vortices and proton fluxtubes in a two-dimensional Ginzburg-Landau model. Within this setting, we compute vortex-fluxtube interaction energies and the outer core analogue of the inner crust pinning energy proxy proposed by \citet{klausner_pinnning_2023}. 
We also study configurations without neutron vortices and confirm the presence of type-1.5-like fluxtube clustering already discussed in \citep{haber2017prd,Wood_2022Univ}, and associate it with a non-monotonic fluxtube-fluxtube effective interaction. 
Vortex-fluxtube tangles, glassy metastability, and far-from-equilibrium line dynamics are outside the present study; see \citep{Drummond2017I,DrummondL2018II,melatos_III_2023}.

\section{Two-condensate model}
\label{sec_model}

We consider a cold neutron star outer core fluid element containing a neutral neutron condensate and a charged proton condensate. The complex scalar fields $\psi_x$ ($x=n,p$ indicates neutrons and protons, respectively) are dimensionless order parameters, normalised so that $|\psi_x|^2=1$ in the uniform phase. 

The fact that both fields $\psi_x$ are scalar already poses a fundamental limitation, since the neutron order parameter is expected to be a tensor field in the outer core \citep{sedrakian2019EPJA}. However, recent simulations by \citet{hattori2026arXiv}, see also \citep{hattori2025}, show that a well-known decomposition of the tensorial neutron order parameter allows numerical studies based on scalar-field minimisation to be extended relatively straightforwardly to this more realistic scenario, by introducing additional complex scalar fields corresponding to the components of the decomposed tensor order parameter.

\subsection{Dimensionless free energy density}

We use a dimensionless two-condensate free energy density closely related to the models of \citet{alford_good_PRB} and \citet{Wood_2022Univ}\footnote{
	We follow the notation of \citep{Wood_2022Univ} and neglect genuine phase-gradient entrainment, i.e., interactions affecting the gradients of the condensate phases~\citep{chamel_review2017JApA,gavassino2020}.
	The precise relation between \eqref{eq_dimensionless_free_energy_density} and \citep{alford_good_PRB,Wood_2022Univ} is given in App.~\ref{app_model_dictionary}. 
}
\begin{align}
	\mathcal{F} ={} & \frac{1}{2}\left(1-|\psi_p|^2\right)^2              
	+\frac{R^2}{2\epsilon}\left(1-|\psi_n|^2\right)^2 \notag \\
	                & +\frac{\alpha}{\epsilon}\left(1-|\psi_p|^2\right)   
	\left(1-|\psi_n|^2\right) \notag \\
	                & +\left|\left(\nabla- i \bm{A}\right)\psi_p\right|^2 
	+\frac{1}{\epsilon}\left|\nabla\psi_n\right|^2 \notag \\
	                & +\kappa^2\left|\nabla\times\bm{A}\right|^2          
	+\frac{g}{\epsilon}\,\nabla|\psi_p|^2\cdot\nabla|\psi_n|^2 .
	\label{eq_dimensionless_free_energy_density}
\end{align}
The model is invariant under local $U(1)_p$ proton gauge transformations and global $U(1)_n$ neutron phase transformations. 
A dimensional interpretation of $\mathcal F$ and the corresponding nondimensionalization are summarized in App.~\ref{app_rescaling}, while the relation of the present model with previous works \citep{alpar1984ApJ,alford_good_PRB,haber2017prd,Wood_2022Univ,Drummond2017I,DrummondL2018II,melatos_III_2023} is detailed in App.~\ref{app_model_dictionary}. 

Following \citep{Wood_2022Univ}, and using the definitions in \eqref{eq_penetration}, the dimensionless parameters in \eqref{eq_dimensionless_free_energy_density} are:
\begin{equation}
	\kappa=\frac{\ell_p}{\xi_p}, \quad
	R=\frac{\xi_p}{\xi_n}, \quad
	\epsilon=\frac{m_n^*n_p^*}{m_p^*n_n^*}, \quad
	\alpha, \quad g.
	\label{eq_dimensionless_parameters}
\end{equation}
Here $\ell_p$ is the proton magnetic penetration length, $\xi_p$ and $\xi_n$ are the proton and neutron coherence lengths, while $m_x^*$ and $n_x^*$ are the condensate masses and homogeneous condensate number densities. For $m_n^*=m_p^*$ and $n_x^*=n_x/2$, as in the normalization of \citet{Wood_2022Univ}, $\epsilon$ reduces to $n_p/n_n$. The spatial coordinates are measured in units of $\xi_p$, so $\nabla$ and $\bm A$ are both dimensionless. 

\subsection{Couplings and repulsion-attraction interpretation}
\label{sec_rep_attr}

The three equivalent free energy densities in \eqref{eq_dimensionless_free_energy_density}, \eqref{eq_dimensional_free_energy_compact}, and \eqref{eq_cicciopazzo} have the same sign interpretation for both the density-density and density-gradient couplings. We therefore discuss it directly in terms of the dimensionless coefficients $\alpha$ and $g$ in~\eqref{eq_dimensionless_free_energy_density}.

The coefficient $\alpha$ controls the local density-density coupling. With the adopted sign convention, overlap of depletions in the condensates (i.e., a region where $|\psi_x|^2<1$ for both $x=n,p$) is energetically favourable when $\alpha<0$, and similarly for an excess in both condensates (i.e., a region where $|\psi_x|^2>1$ for both $x=n,p$). For this reason, we refer to the case $\alpha>0$ ($\alpha<0$) as ``repulsive'' (``attractive''), in a purely local sense.

Interestingly, this local interpretation of the sign of $\alpha$ is also typically reflected in the effective non-local attraction or repulsion between a vortex and a fluxtube~\citep{Drummond2017I,Shukla_PRD_2024}, since $\alpha<0$ favours overlap of depleted neutron and proton regions. This non-local interaction is, however, an emergent property of the full  configuration and does not follow from the local coupling term alone. In particular, local attraction does not by itself guarantee vortex-fluxtube pinning, as other contributions may be present.

We extend the same local sign interpretation to the coefficient $g$. Since the density gradients of overlapping vortex and fluxtube cores are approximately aligned, for $g<0$ we expect that superimposing the defects is locally more energetically favourable than in the repulsive case~$g>0$. As for $\alpha$, whether this results in an effective non-local attraction between the defects depends on the full free energy \eqref{eq_free_energy} of the whole configuration.

\section{Free energy minimisation}
\label{sec_numerical_procedure}

We consider a two-dimensional cross-section in which neutron vortices and proton fluxtubes are locally aligned along an common direction, chosen as the $z$-axis. 
The total free energy per unit length is
\begin{equation}
	F[\psi_p,\psi_n,\bm A]
	=
	\int_{\mathcal D}\mathcal F\,dx\,dy,
	\label{eq_free_energy}
\end{equation}
where $\mathcal D$ is a rectangle of size $L_x\times L_y$ in the dimensionless coordinates $(x,y)$, corresponding to a physical area $L_xL_y\xi_p^2$. We use centred coordinates with bounds $|x| \leq L_x/2$ and~$|y| \leq L_y/2$.

We use two minimisation procedures. In the first, QPBC impose the winding numbers $\mathcal Q_n$ and $\mathcal Q_p$, while the fields and defect positions are allowed to relax. In the second, the condensate phases are prescribed to keep the defects at fixed positions and determine the free energy as a function of their separation.

\subsection{Full minimisation with QPBC}
\label{sec_qpbc}

For configurations with net vorticity or magnetic flux, ordinary periodic boundary conditions cannot be imposed on the condensate phases. We therefore use QPBC, see e.g. \citep{doria_1988,Wood2019PhRvB,Doran2020,Magistrelli_2026pra}, to select a prescribed topological sector.

We denote by $\mathcal Q_n$ and $\mathcal Q_p$ the neutron and proton winding numbers in the cell. In the proton sector, $\mathcal Q_p$ is also the number of flux quanta, so that
\begin{equation}
	\int_{\mathcal D}B_z\,dx\,dy
	=
	\langle B\rangle L_xL_y
	=
	2\pi\mathcal Q_p.
	\label{eq_fluxQp}
\end{equation}
We thus write
\begin{equation}
	\bm A
	=
	\frac{\langle B\rangle}{2}\,\hat{\bm z}\times\bm x
	+
	\delta\bm A,
\end{equation}
where $\delta\bm A$ is periodic over $\mathcal{D}$, so we can use standard periodic boundary conditions for it. 

The proton field satisfies the two-dimensional QPBC of~\citet{doria_1988}
\begin{align}
	\psi_p(x+L_x,y)
	  & = 
	e^{i\chi_x^p(y)}\psi_p(x,y),
	\\
	\psi_p(x,y+L_y)
	  & = 
	e^{i\chi_y^p(x)}\psi_p(x,y),
\end{align}
with
\begin{equation}
	\label{eq_struzzo_p}
	\chi_x^p(y)
	=
	\frac{\langle B\rangle L_x}{2}y,
	\qquad
	\chi_y^p(x)
	=
	-\frac{\langle B\rangle L_y}{2}x.
\end{equation}
For the neutron field, we impose
\begin{align}
	\psi_n(x+L_x,y)
	  & = 
	e^{i\chi_x^n(y)}\psi_n(x,y),
	\\
	\psi_n(x,y+L_y)
	  & = 
	e^{i\chi_y^n(x)}\psi_n(x,y),
\end{align}
where
\label{eq_struzzo_n}
\begin{equation}
	\chi_x^n(y)
	=
	\frac{\pi\mathcal Q_n}{L_y}y,
	\qquad
	\chi_y^n(x)
	=
	-\frac{\pi\mathcal Q_n}{L_x}x.
\end{equation}
Thus, $|\psi_n|$, $|\psi_p|$, and $B_z$ are periodic over $\mathcal D$, while the prescribed phase twists impose $\mathcal Q_n$ and $\mathcal Q_p$.

In these minimisations, the real and imaginary parts of both condensates and the periodic field $\delta\bm A$ are varied, and the topological defects arrange themselves in the most energetically favourable positions. 
The QPBC determine also the average offset of the corresponding defect array~\citep{Wood2019PhRvB}; with the conventions adopted in \eqref{eq_struzzo_p} and \eqref{eq_struzzo_n} and our centered coordinates, an isolated defect is centred in the computational cell.

The QPBC used here are not intended to reproduce realistic values of the average vorticity and thermodynamic magnetic field expected in pulsars or magnetars, which are set by the two ratios $\mathcal{Q}_x/(L_x L_y)$, $x=n,p$. Here the QPBC are simply a device to restrict the field configurations to a given topological sector. Unavoidable finite-size effects can only be kept under control by probing different $L_x$ and $L_y$ at fixed topological charges~$\mathcal{Q}_x$.

\subsection{Amplitude minimisation with phase imprinting and pinning energy}
\label{sec_phase_imprinting}

To determine the interaction energy at a prescribed separation $d$, we fix the phases of $\psi_n$ and $\psi_p$ so that their singularities remain at specified positions, and minimise over the amplitudes $|\psi_n|$, $|\psi_p|$, and the vector potential.

The imprinted phases include image defects in repetitions of $\mathcal D$ around the central cell. Images with the same winding approximate the periodic phase field, while alternating windings in a checkerboard pattern approximate hard-wall boundaries~\citep{Wood2019PhRvB,Doran2020,Magistrelli_2026pra,saffman1995vortex}. We use both prescriptions to check finite-size effects.

Let $F_{\rm min}(d;\alpha,g)$ be the constrained minimum at separation $d$. Even for $\alpha=g=0$, this quantity can depend on $d$ because of the finite domain and the prescribed images. We subtract this geometrical contribution at the same separation,
\begin{equation}
	\Delta E(d;\alpha,g)
	=
	F_{\rm min}(d;\alpha,g)
	-
	F_{\rm min}(d;0,0).
	\label{eq_DeltaF_fixed_d}
\end{equation}
We take
\begin{equation}
	d_{\rm ref}=L_x/2
\end{equation}
and define
\begin{equation}
	E_{\rm pin}(d;\alpha,g)
	=
	\Delta E(d;\alpha,g)
	-
	\Delta E(d_{\rm ref};\alpha,g).
	\label{eq_Epin_definition}
\end{equation}
Hence $E_{\rm pin}(d_{\rm ref};\alpha,g)=0$, while $E_{\rm pin}<0$ indicates that separation $d$ is favoured with respect to the reference configuration. If the minimum occurs at $d=d_{\rm min}$, the corresponding unpinning energy is $-E_{\rm pin}(d_{\rm min};\alpha,g)>0$.

The associated force per unit length is
\begin{equation}
	f_{\rm pin}(d)
	=
	-E'_{\rm pin}(d)
	=
	-\Delta E'(d).
	\label{eq_pinning_force_landscape}
\end{equation}
We check the stability of $E_{\rm pin}$ by varying $L_x$, $L_y$, and the image prescription.

\subsection{Details of the minimisation procedure}
\label{sec_details_min}

We discretise $\mathcal D$ on a $N_x\times N_y$ grid. The results shown below use approximately $N_{x,y}=5L_{x,y}$. The condensate fields are stored on nodes, $A_x$ and $A_y$ on links, and $B_z=\partial_xA_y-\partial_yA_x$ on plaquette centres.

The boundary conditions are implemented with ghost cells. The discrete free energy is minimised with the full-batch Adam optimiser in PyTorch~\citep{pytorch_2019arXi}. In the QPBC calculations, the trainable variables are the real and imaginary parts of $\psi_n$ and $\psi_p$ and the periodic components of $\delta\bm A$. In the phase-imprinting calculations, the phases are fixed and the amplitudes and vector potential are varied.

\section{Numerical results}
\label{sec_results}

For realistic applications to the outer core of a neutron star, a microscopic calibration of the $\{ \kappa , R , \epsilon , \alpha , g \}$ parameters is necessary: these parameters should be related consistently to the baryon density, assuming beta equilibrium. Although some of the parameters can be mapped onto available microscopic nuclear information (e.g., $\epsilon=m_n^*n_p^*/(m_p^*n_n^*)$, which reduces to $n_p/n_n$ under the equal-mass normalization of \citet{Wood_2022Univ}), such a calibration of the phenomenological model in \eqref{eq_dimensionless_free_energy_density} is not presently fully available. On the other hand, in the absence of this calibration, a full parameter study involving all five $\{ \kappa , R , \epsilon , \alpha , g \}$ is impractical, and thus not the focus of the present work. We therefore fix
\begin{equation}
	\label{eq_params} 
	\kappa=1.44,\quad R=0.4, \quad \epsilon=0.1 ,  
\end{equation}
to the fiducial values of \citet{Wood_2022Univ}, and explore the effect of the two couplings $\alpha$ and $g$. The corresponding coupling sector coincides with that of \citet{alford_good_PRB}; see App.~\ref{app_model_dictionary} for the explicit comparison.

For the fiducial values in \eqref{eq_params}, one unit in the dimensionless coordinates $x$ and $y$ corresponds to~$\xi_p \approx 31\,{\rm fm}$~\citep{Wood_2022Univ}.

\subsection{Stability domain}
\label{sec_stability}

We first estimate a conservative range of $\alpha$ and $g$ for which the homogeneous state and the near-homogeneous gradient sector are locally stable. All numerical experiments below, including those that explore states far from the homogeneous case, are thus performed well within this working window.

The simplest stability requirement is just the usual positive definiteness condition for the homogeneous sector, see e.g. \citep{pethick_book_2008,KobyakovSupercond}. Defining
\begin{equation}
	u_x=|\psi_x|^2, \quad u_x=1+\delta u_x,  \quad (x=n,p),
\end{equation}
for homogeneous configurations we have 
\begin{equation}
	\mathcal F 
	=
	\frac12(\delta u_p)^2
	+
	\frac{R^2}{2\epsilon}(\delta u_n)^2
	+
	\frac{\alpha}{\epsilon}\delta u_p\delta u_n \, .
\end{equation}
Since $\epsilon>0$, positive definiteness of the above form therefore gives the stability requirement in \citep{Wood_2022Univ},
\begin{equation}
	\label{eq_alphastable}
	|\alpha|<R\sqrt{\epsilon}.
\end{equation}
Here, we extend the same reasoning to the gradient sector, which gives an analogous estimate for $g$. Setting $\alpha = \mathbf A = 0$ and writing $\psi_x=\rho_x \exp{(i\theta_x)}$, one finds
\begin{multline}
	\mathcal F
	=
	|\nabla\rho_p|^2+\rho_p^2|\nabla\theta_p|^2
	+
	\frac{1}{\epsilon}\left(|\nabla\rho_n|^2+\rho_n^2|\nabla\theta_n|^2\right)
	\\+
	\frac{4g}{\epsilon}\rho_p\rho_n\nabla\rho_p\cdot\nabla\rho_n.
\end{multline}
The phase-gradient terms $|\nabla\theta_x|^2$ are positive independently of $g$, meaning that possible instability is only in the density-gradient quadratic form,
\begin{equation}
	|\nabla\rho_p|^2
	+
	\frac{1}{\epsilon}|\nabla\rho_n|^2
	+
	\frac{4g}{\epsilon}\rho_p\rho_n\nabla\rho_p\cdot\nabla\rho_n \, ,
\end{equation}
whose pointwise positivity gives
\begin{equation}
	|g|<\frac{\sqrt{\epsilon}}{2\rho_p\rho_n}
	=
	\frac{ 1}{2 }
	\sqrt{\frac{\epsilon}{  u_p u_n }} .
\end{equation}
For configurations close to the homogeneous state, $u_p=u_n=1$, this reduces to
\begin{equation}
	\label{eq_gstable}
	|g|< \sqrt{\epsilon}/2 \, .
\end{equation}
Interestingly, the above equation tells us that, if one demands positivity for arbitrary gradients, no nonzero constant $g$ satisfies the pointwise condition globally.

For the parameters in \eqref{eq_params}, the analytic estimates \eqref{eq_alphastable} and \eqref{eq_gstable} give
\begin{equation}
	\label{eq_bounds}
	|\alpha|<0.1265 \, ,
	\qquad
	|g|<0.1581 .
\end{equation}
These are local analytic estimates and do not establish global boundedness of the full functional for general configurations.

For this reason, and to stress-test our PyTorch optimiser, we also perform numerical convergence scans. First, we vary one coupling at a time, with the other set to zero, for homogeneous configurations and for configurations containing either a single fluxtube ($|\mathcal Q_p| =1  $, $\mathcal Q_n =0 $) or a single vortex ($|\mathcal Q_n| =1  $, $\mathcal Q_p =0 $) imposed through the QPBC. 
This gives the limits reported in Tab.~\ref{tab_single_coupling_stability}, beyond which the minimisation procedure does not converge. The numerical limits are consistent with the analytic estimates \eqref{eq_bounds} up to the scan resolution, and the presence of a topological defect changes them only mildly, at the level of a few~$10^{-3}$.

We then vary both couplings simultaneously and record representative boundary points of the numerical convergence region in the defect sectors, as reported in Tab.~\ref{tab_two_coupling_stability}. The region is only weakly deformed when both couplings are active. We therefore use \eqref{eq_bounds} as a conservative estimate of the working domain and restrict to $|\alpha|\le 0.1$ and~$|g|\le 0.1$.

\begin{table}
	\centering
	\begin{tabular}{lccc}
		\hline\hline
		                   & fluxtube   & Vortex     & Homogeneous \\
		\hline
		$\alpha_{\rm min}$ & $-0.12632$ & $-0.12755$ & $-0.12632$  \\
		$\alpha_{\rm max}$ & $ 0.12632$ & $ 0.12632$ & $ 0.12632$  \\
		$g_{\rm min}$      & $-0.15979$ & $-0.16102$ & $-0.16020$  \\
		$g_{\rm max}$      & $ 0.16020$ & $ 0.16142$ & $ 0.16020$  \\
		\hline\hline
	\end{tabular}
	\caption{Numerical convergence limits obtained by varying one coupling parameter while the other is set to zero, to be compared with the local estimates \eqref{eq_bounds}. The scan is performed with configurations containing no defects (homogeneous), one fluxtube or one vortex, with a resolution of $0.00041$ for both parameters.}
	\label{tab_single_coupling_stability}
\end{table}

\begin{table}
	\centering
	\begin{tabular}{cc}
		\hline\hline
		Fluxtube $(g,\alpha)$ & Vortex $(g,\alpha)$ \\
		\hline
		$(-0.1557,-0.1254)$   & $(-0.1604,-0.1275)$ \\
		$( 0.1616,-0.1254)$   & $( 0.1683,-0.1275)$ \\
		$( 0.1557, 0.1254)$   & $( 0.1560, 0.1254)$ \\
		$(-0.1616, 0.1254)$   & $(-0.1604, 0.1254)$ \\
		\hline\hline
	\end{tabular}
	\caption{Representative boundary points of the numerical convergence region when both couplings are varied in a domain containing a fluxtube or a vortex.}
	\label{tab_two_coupling_stability}
\end{table}

\subsection{Coupling-induced deformation of single defects}
\label{sec_profiles}

As a first check of the minimisation procedure (Sec.~\ref{sec_details_min}) and QPBC implementation (Sec.~\ref{sec_qpbc}), we study single-defect profiles for different coupling parameters. We compare a single vortex $(\mathcal Q_n=\pm1, \mathcal Q_p=0)$ and a single fluxtube $(\mathcal Q_n=0, \mathcal Q_p=\pm1)$ with their uncoupled profiles. Coupling-induced deformations of isolated fluxtubes were already discussed by \citet{alford_good_PRB}; here we recover this general behaviour and also consider the complementary case of an isolated neutron vortex.

\begin{figure}
	\centering
	\includegraphics[width=\linewidth]{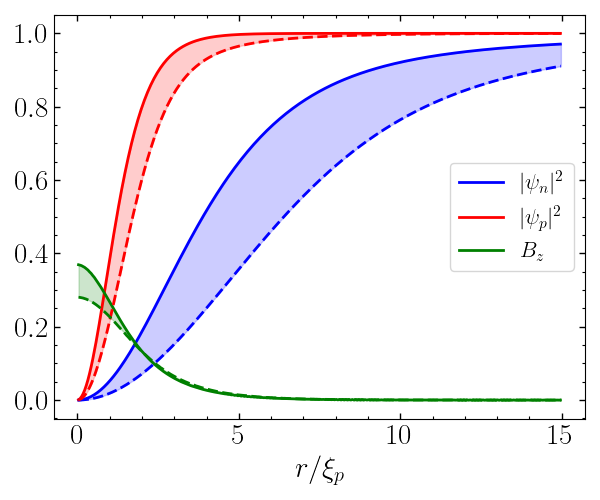}
	\caption{
		Single-defect profiles in a domain with $L_x=L_y=50$. Solid curves show the uncoupled case. Dashed curves show the coupled case $\alpha=g=0.1$, which gives the largest deformation in the scan. The shaded bands indicate the spread over the explored coupling values. The $|\psi_n|^2$ profile refers to a single vortex $(|\mathcal Q_n|=1, \mathcal Q_p=0)$, while $|\psi_p|^2$ and $B_z$ refer to a single fluxtube $(\mathcal Q_n=0, \mathcal Q_p=1)$. The case $(\mathcal Q_n=0, \mathcal Q_p=-1)$ is identical, but with $B_z$ reversed.
	}
	\label{fig:profiles_no_int}
\end{figure}

\begin{figure*}
	\centering
	\begin{subfigure}{0.32\textwidth}
		\centering
		\includegraphics[width=\linewidth]{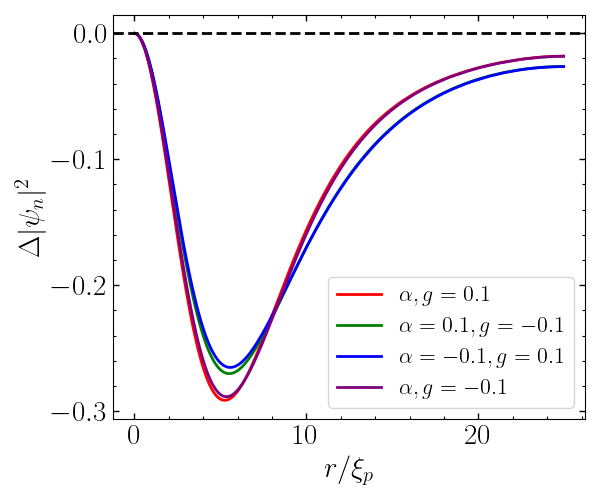}
		\caption{Dimensionless neutron condensate amplitude~$(\mathcal Q_n=1, \mathcal Q_p=0)$.}
		\label{fig:delta_psin_int}
	\end{subfigure}
	\hfill
	\begin{subfigure}{0.32\textwidth}
		\centering
		\includegraphics[width=\linewidth]{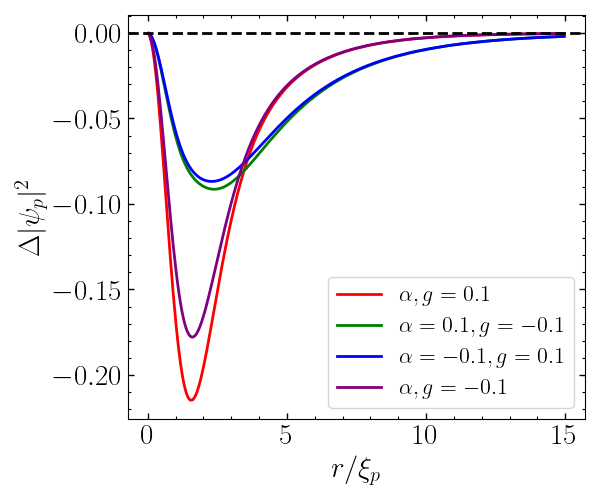}
		\caption{Dimensionless proton condensate amplitude~$(\mathcal Q_n=0, \mathcal Q_p=1)$.}
		\label{fig:delta_psip_int}
	\end{subfigure}
	\hfill
	\begin{subfigure}{0.32\textwidth}
		\centering
		\includegraphics[width=\linewidth]{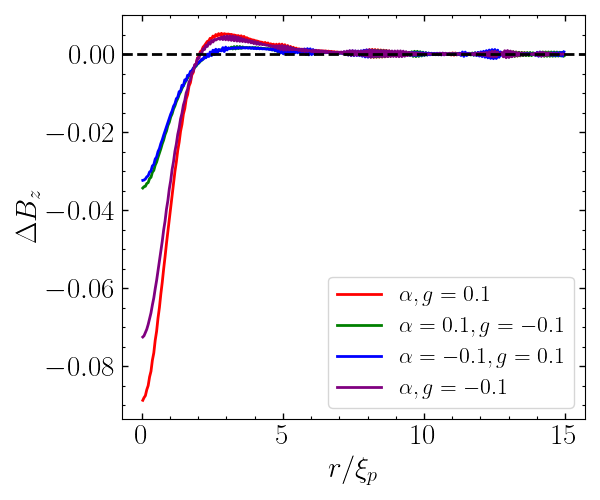}
		\caption{Dimensionless magnetic field profile of a fluxtube~$(\mathcal Q_n=0, \mathcal Q_p=1)$.}
		\label{fig:delta_bz_int}
	\end{subfigure}
	\caption{Differences in the relaxed single-defect radial profiles induced by inter-condensate couplings $\alpha,g= \pm 0.1$ with respect to the case~$\alpha =g=0$.}
	\label{fig:profiles_with_int}
\end{figure*}

Fig.~\ref{fig:profiles_no_int} compares the uncoupled profiles ($\alpha =g=0$, solid lines) with the coupled case $\alpha=g=0.1$, which gives the largest deformation from the uncoupled profiles within the explored range while remaining within the window established in Sec.~\ref{sec_stability}. The intermediate coupled cases lie within the shaded bands. The couplings modify the effective core sizes and the magnetic-field profile, while the changes remain moderate even for values close to the edge of our working range. Fig.~\ref{fig:profiles_with_int} shows the changes for the four combinations $\alpha,g=\pm0.1$ relative to~$\alpha=g=0$.

These results show that the local couplings do not radically change the structure of isolated topological defects within the parameter range considered here. This provides a useful baseline for the following sections: despite the moderate deformation of isolated defects, the same local couplings can have a much stronger effect on the energetics when vortex and fluxtube cores partially or fully overlap, with a noticeable effect up to distances of several~$\xi_p$.

\subsection{Vortex-fluxtube pinning}
\label{sec_pinning}

\begin{figure}
	\centering
	\includegraphics[width=\linewidth]{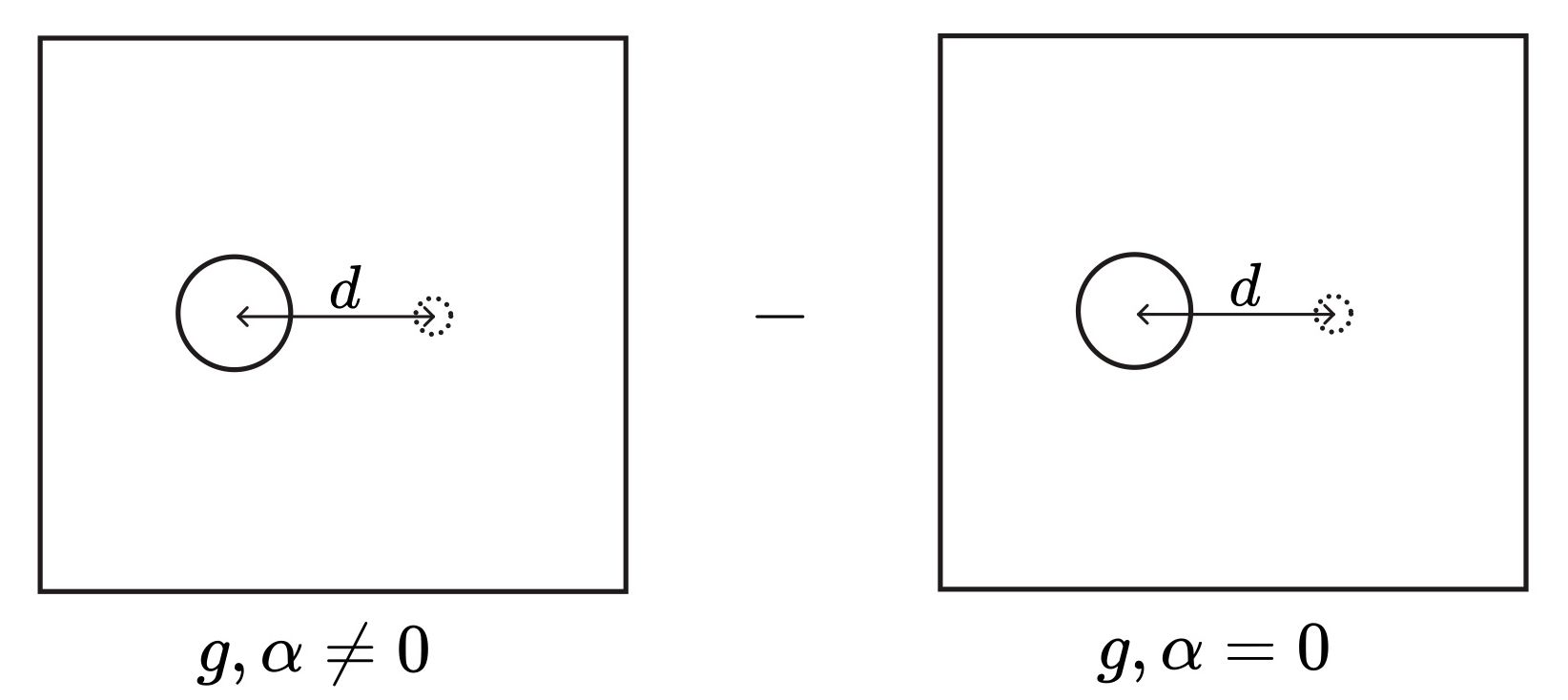}
	\caption{Schematic constrained minimization protocol used to extract the vortex-fluxtube interaction energy at fixed separation $d$, see Sec.~\ref{sec_phase_imprinting}: each box represents $\mathcal D$ with $|\mathcal{Q}_n| = |\mathcal{Q}_p| = 1$; the wider cylinder denotes the vortex and the narrower cylinder denotes the proton fluxtube. 
		The difference between the two configurations represents~\eqref{eq_DeltaF_fixed_d}.
	}
	\label{fig_interaction_schematic}
\end{figure}

We next consider a neutron vortex and a proton fluxtube at fixed separation $d$, and follow the procedure outlined in Sec.~\ref{sec_phase_imprinting}, which is illustrated in Fig.~\ref{fig_interaction_schematic}.

\begin{figure}
	\centering
	\includegraphics[width=\linewidth]{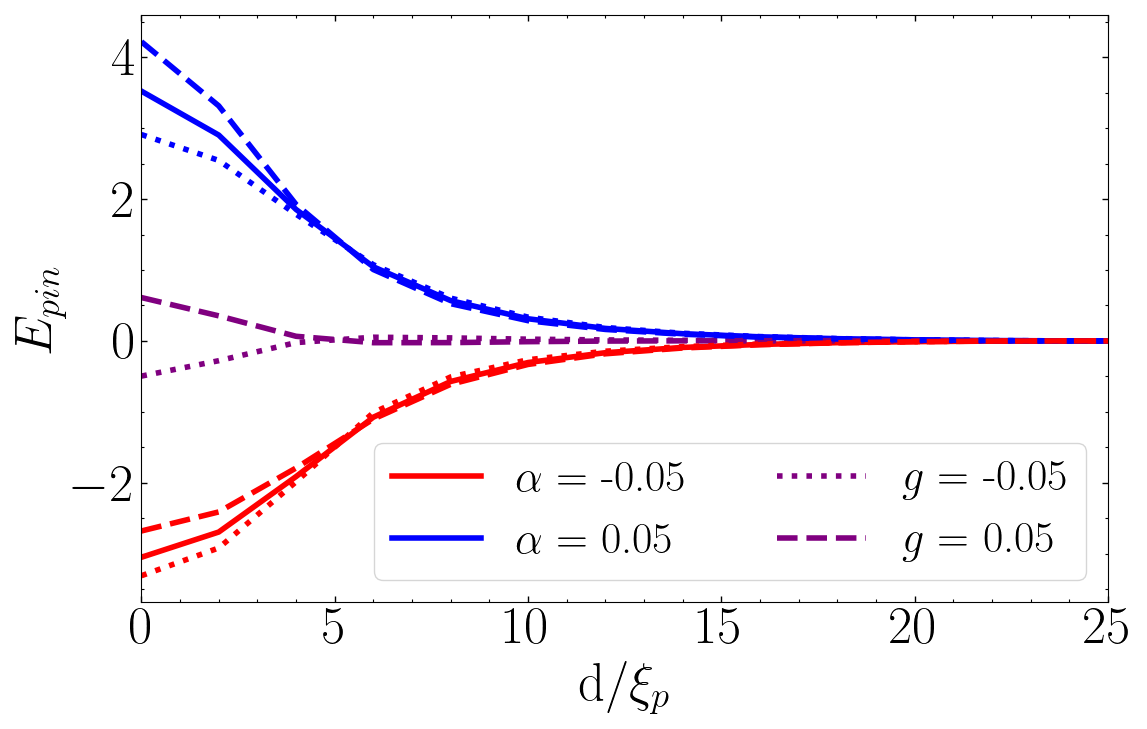}
	\caption{The vortex-fluxtube pinning energy  per unit length $E_{\rm pin}(d;\alpha,g)$ as a function of their imposed separation $d$. The effects of $\alpha$ and $g$ are essentially additive. The same result is obtained for the anti-aligned case $\mathcal{Q}_n=-\mathcal{Q}_p=1$, see Fig.~\ref{fig_vortex_fluxtube_diff}. For the fiducial parameters used here, one dimensionless unit of $E_{\rm pin}$ corresponds to~$0.13\,{\rm MeV\,fm^{-1}}$. 
	}
	\label{fig_vortex_fluxtube_energy_distance}
\end{figure}

Figure~\ref{fig_vortex_fluxtube_energy_distance} shows the interaction energy as the vortex and fluxtube are separated from an initially overlapping configuration: at each scanned $d$ we find the optimal configuration with $\mathcal{Q}_n=\mathcal{Q}_p=1$ and extract the pinning energy per unit length in \eqref{eq_Epin_definition}.

For attractive couplings ($\alpha<0$ and/or $g<0$, see Sec.~\ref{sec_rep_attr}), the energy $E_{\rm pin}(d;\alpha,g)$ increases with separation. The overlapped state at $d=0$ is therefore energetically preferred and indicates vortex-fluxtube pinning. For repulsive local couplings the overlapped state is disfavored, as expected.

In our finite-domain setting, the energy landscape as a function of $d$ is affected by periodic images for $d\gtrsim L_x/2$, an unavoidable limitation of finite-domain constrained minimization. When periodic images mimicking QPBC are used, the curves in Fig.~\ref{fig_vortex_fluxtube_energy_distance} are exactly symmetrical around their zeroes at $d=d_{\rm ref}$. However, the displayed part $0<d<d_{\rm ref}$ is unchanged with respect to the hard-wall case for sufficiently large domains; in this case, $L_x=L_y=50$ is more than enough to obtain robust results. This provides a check that the extracted $E_{\rm pin}(d;\alpha,g)$ is accurate up to the plateau; in the bulk limit, it would simply remain flat.

\begin{figure}
	\centering
	\includegraphics[width=\linewidth]{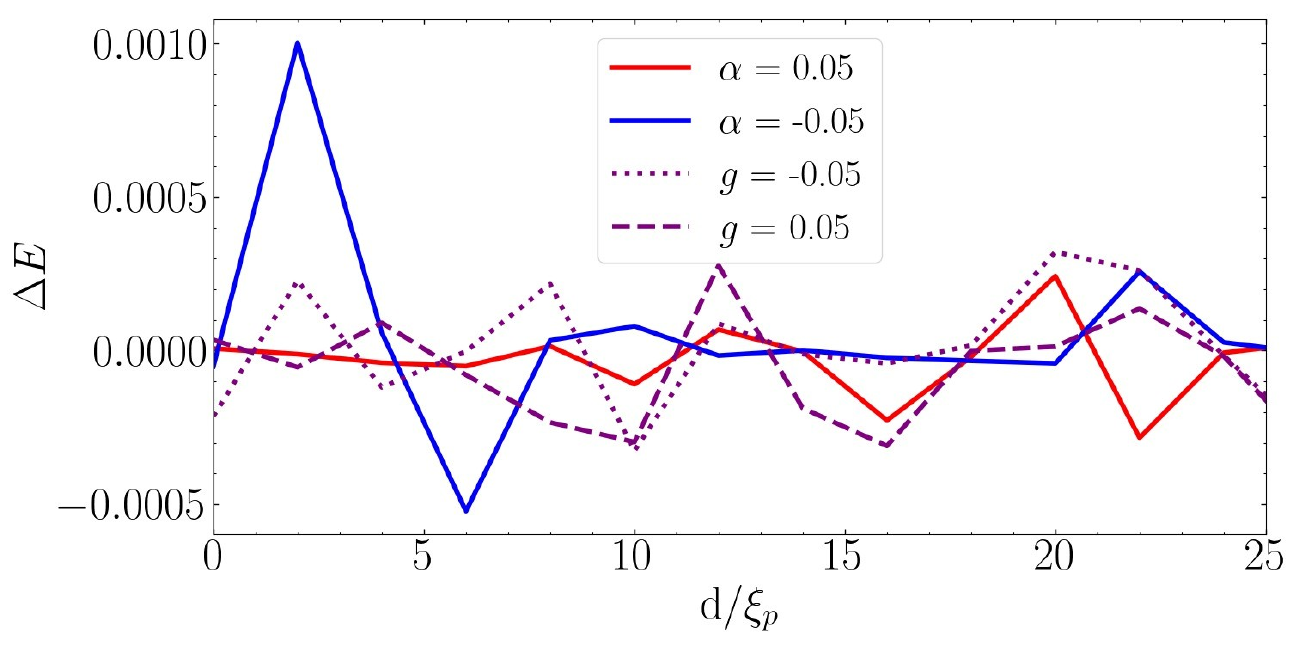}
	\caption{The difference of $E_{\rm pin}(d;\alpha,g)$ between the aligned case $\mathcal{Q}_n=\mathcal{Q}_p=1$ and the anti-aligned case $\mathcal{Q}_n=-\mathcal{Q}_p=1$. The difference is absolute, to be compared against the curves in Fig.~\ref{fig_vortex_fluxtube_energy_distance} and essentially due to the numerical resolution of the minimisation procedure~(the peak at 0.001 of the blue curve corresponds to a relative error of about~$0.03\%$).
	}
	\label{fig_vortex_fluxtube_diff}
\end{figure}

A final point is worth commenting. The experiment in Fig.~\ref{fig_vortex_fluxtube_energy_distance} considers an aligned vortex and fluxtube, namely $\mathcal{Q}_n=\mathcal{Q}_p=1$ (by symmetry, the exact same result would be obtained for the other aligned case $\mathcal{Q}_n=\mathcal{Q}_p=-1$). However, it may not be immediately obvious whether the same result would hold for the anti-aligned case  $\mathcal{Q}_n=-\mathcal{Q}_p=1$, or $-\mathcal{Q}_n=\mathcal{Q}_p=1$.
The answer is yes, since in our model there is no entrainment, namely no coupling between the neutron and proton phase gradients. As a result, the phase sector is additive, and the velocity-field contribution cancels when equivalent reference configurations are subtracted, making the extracted $E_{\rm pin}(d;\alpha,g)$ controlled by the core overlap, magnetic profile, and density-gradient terms. This analytical argument is checked numerically in Fig.~\ref{fig_vortex_fluxtube_diff}.

\subsection{Asymptotic binding energy}
\label{sec_asymptotic_proxy}

Finite boxes cannot directly realize infinite vortex-fluxtube separation, so our $d_{\rm ref}$ is at most $\sqrt{L_x^2+L_y^2}/2$ if the defects are placed on the diagonal of $\mathcal D$. 
This does not appear to be a problem as long as $d_{\rm ref}$ lies well within the range of distances where the pinning energy has reached a plateau, as shown in Fig.~\ref{fig_vortex_fluxtube_energy_distance}. However, our procedure is based on energy differences between configurations that all have fixed~$|\mathcal{Q}_n|=|\mathcal{Q}_p|=1$.

A different estimate of the pinning energy was proposed by \citet{klausner_pinnning_2023}, based on configurations containing different numbers of defects. They refer to this quantity as a ``binding energy'', and use it as a proxy for vortex-nucleus pinning in the inner crust. In particular, their proxy estimates the energy cost of moving a vortex from a configuration in which it overlaps with a nucleus ($d=0$) to one in which the two are infinitely separated.

\begin{figure*}
	\centering
	\includegraphics[width=0.8\textwidth]{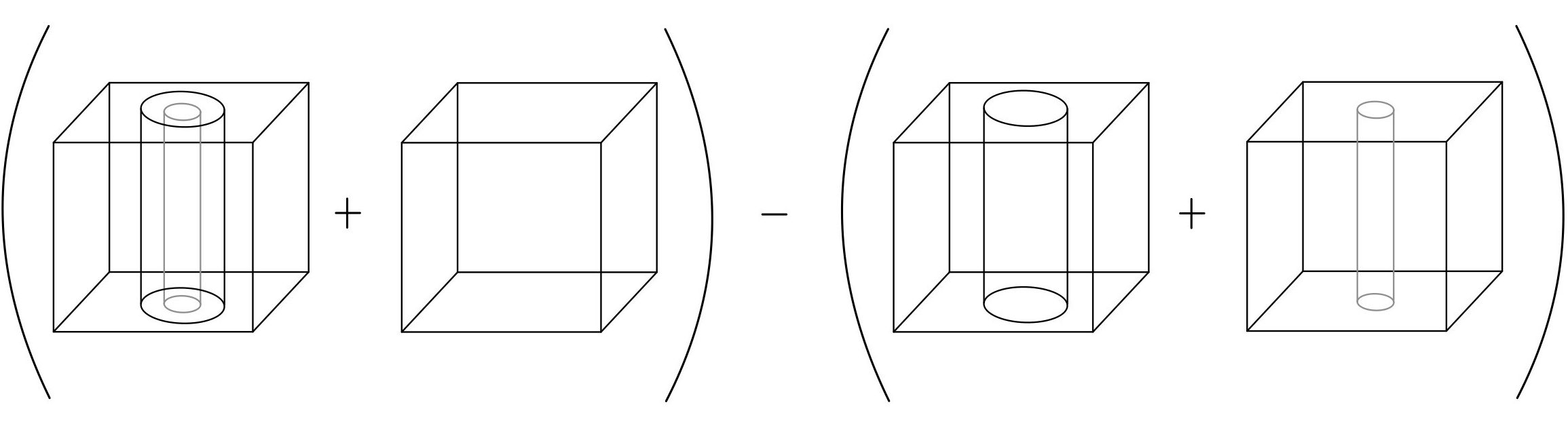}
	\caption{Schematic energy-difference construction used as an asymptotic proxy for pinning. All four configurations are evaluated at the same $\alpha$ and $g$ but have different $\mathcal{Q}_n$ and $\mathcal{Q}_p$. The wider cylinder denotes a neutron vortex and the thinner cylinder denotes a proton fluxtube.}
	\label{fig_klausner_proxy}
\end{figure*}

It is therefore interesting to check whether the two approaches give the same result, but we must first adapt the vortex-nucleus procedure of~\citet{klausner_pinnning_2023} to our case of vortex-fluxtube pinning: a schematic representation is shown in Fig.~\ref{fig_klausner_proxy}, cf.~Fig.~1 in~\citep{klausner_pinnning_2023} and~Fig.~1 in~\citep{donati2004}. This would also validate their binding-energy procedure as a proxy for pinning.

Unlike~\eqref{eq_DeltaF_fixed_d}, the proxy compares four minimized configurations all having the same $\mathcal{F}$ parameters, in particular $\alpha$ and $g$, and forms an energy difference designed to isolate the binding energy, irrespective of the details of the energy landscape at finite distances. Another important difference is that the configurations used in this proxy do not all have~$|\mathcal{Q}_n|=|\mathcal{Q}_p|=1$.

\begin{figure}
	\centering
	\includegraphics[width=\linewidth]{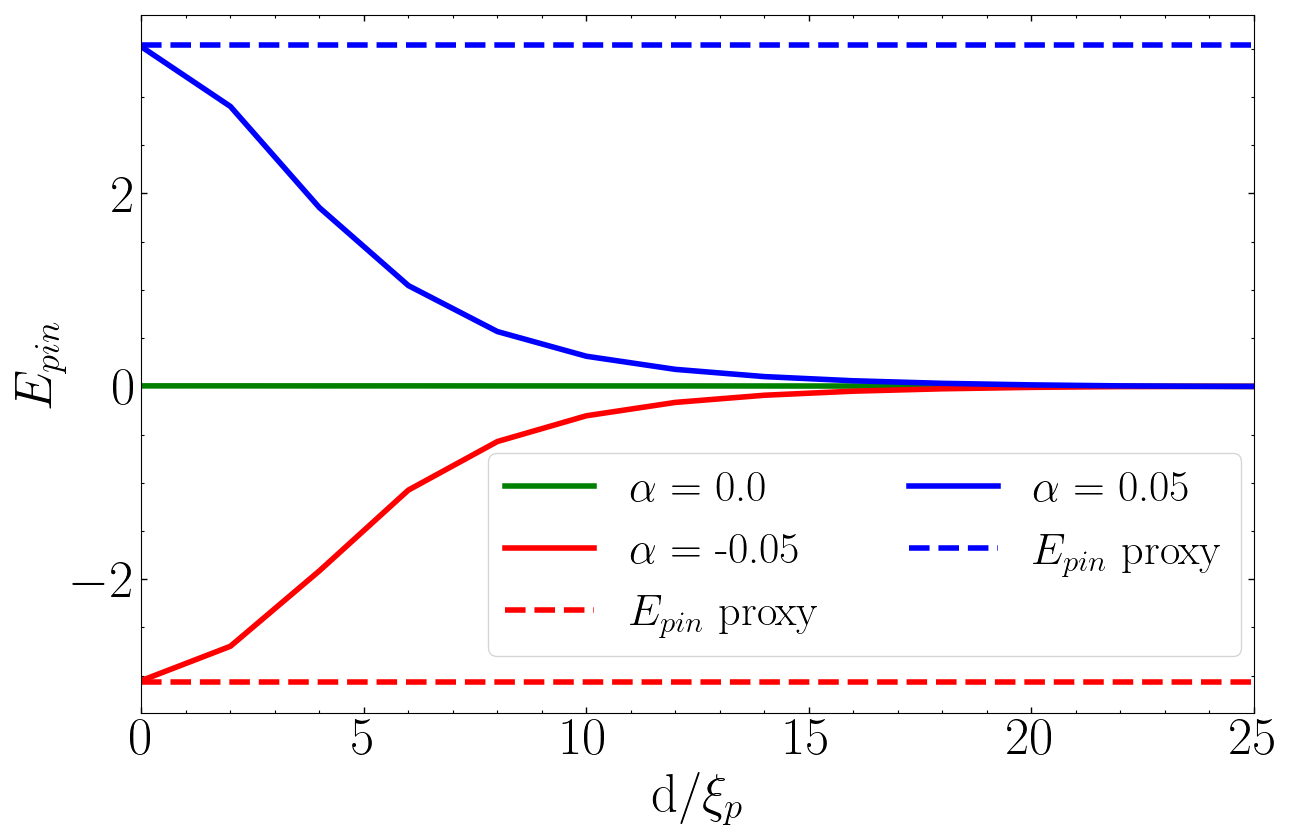}
	\caption{Comparison between direct separation-dependent interaction energies and asymptotic proxy estimates. Horizontal lines indicate proxy values for representative choices of $\alpha$ and~$g=0$. }
	\label{fig_proxy_vs_direct}
\end{figure}

Fig.~\ref{fig_proxy_vs_direct} compares the proxy with the direct constrained-minimization result. For the representative cases shown with $g=0$, as well as more generally for nonzero $g$, a separation $d\approx 20$ is already enough to reproduce the asymptotic regime within the finite computational domain. Although it may not be immediately obvious from the scheme in Fig.~\ref{fig_klausner_proxy}, the proxy also gives zero binding energy when $\alpha=g=0$ exactly by construction. The same conversion applies to the proxy energies, so one unit of dimensionless binding energy per unit length corresponds to~$0.131\,{\rm MeV\,fm^{-1}}$, see~\eqref{eq_facocero}.

\begin{figure}
	\centering
	\includegraphics[width=\linewidth]{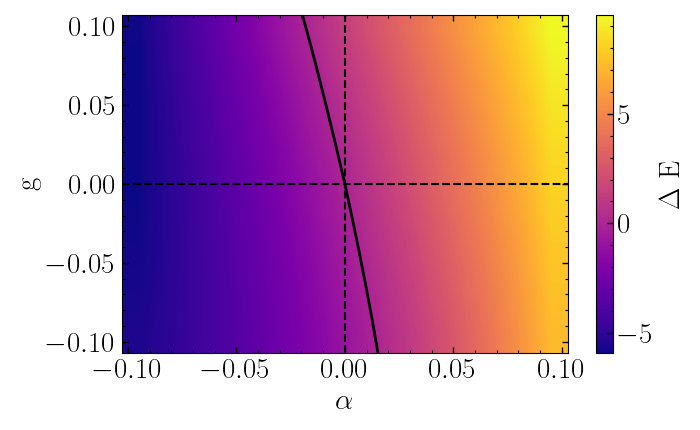}
	\caption{Binding energy as a function of $\alpha$ and $g$. Dashed lines show $\alpha=0$ and $g=0$. The solid contour marks the zero of the binding energy (the negative values on the left part of the colour map are a proxy for attractive vortex-fluxtube pinning).}
	\label{fig_pinning_heatmap}
\end{figure}

The full dependence of the binding energy per unit length on $\alpha$ and $g$ is shown in Fig.~\ref{fig_pinning_heatmap}. The dashed lines are just to guide the eye and indicate $\alpha=0$ and $g=0$, while the solid contour line indicates the zero of the binding energy and thus separates attractive and repulsive regimes. The main contribution to the sign of the binding energy is set by $\alpha$, while $g$ provides a secondary correction.

\subsection{Fluxtube bouquets}
\label{sec_bouquets}

The vortex-fluxtube pair analysis of Sec.~\ref{sec_pinning} shows that attractive couplings also favour vortex-fluxtube overlap. However, in a neutron star outer core, the fluxtube density can exceed the vortex density by many orders of magnitude. 
Hence, a local equilibrium can be set by competition between fluxtube-fluxtube repulsion and vortex-fluxtube attraction, a situation that cannot be probed with our previous procedure involving only configurations with~$|\mathcal{Q}_n|=|\mathcal{Q}_p|=1$.

We therefore now explore configurations in which a neutron vortex can bind several nearby fluxtubes. Thinking of the fluxtubes as flowers and of the vortex as the wrapping that holds them together, we call these composite states, in which a vortex is decorated by several fluxtubes, ``bouquets''.

\begin{figure*}
	\centering
	\begin{subfigure}{0.48\textwidth}
		\centering
		\includegraphics[width=\linewidth]{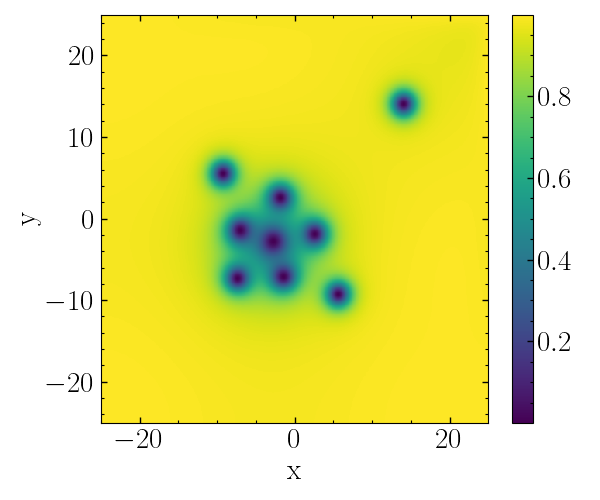}
		\caption{Proton condensate amplitude~$|\psi_p|^2$.}
		\label{fig:bouquet_p}
	\end{subfigure}
	\hfill
	\begin{subfigure}{0.48\textwidth}
		\centering
		\includegraphics[width=\linewidth]{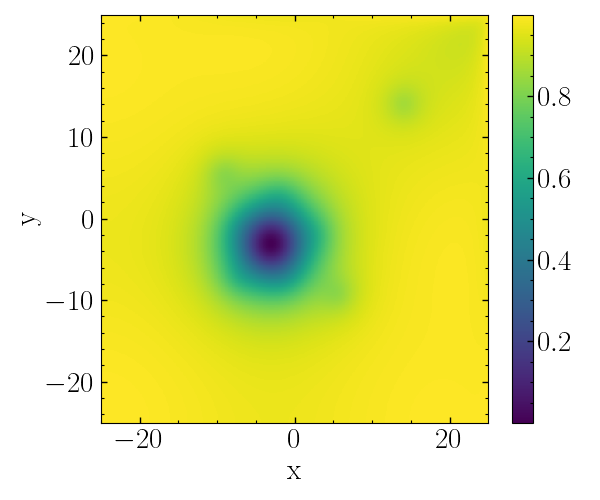}
		\caption{Neutron condensate amplitude~$|\psi_n|^2$.}
		\label{fig:bouquet_n}
	\end{subfigure}
	\caption{Formation of a fluxtube bouquet in a cell with QPBC with $\mathcal Q_n=1$ and $\mathcal Q_p =9$, at $\alpha=-0.05, g=0$. Several proton fluxtubes are locally bound to the neighborhood of a neutron vortex.}
	\label{fig:bouquet}
\end{figure*}

Figure~\ref{fig:bouquet} shows an example with $\mathcal Q_n=1$ and $\mathcal Q_p =9$, where a bouquet forms for $\alpha=-0.05$: the proton-amplitude panel shows several fluxtube cores gathered around the vortex, while the neutron-amplitude panel identifies the vortex location. Some fluxtubes remain farther from the core of the bouquet. It is difficult to determine whether this is due to the very slow convergence of the minimisation procedure, so that the shown configuration has not yet reached the true minimum, or whether the vortex has effectively exhausted its ability to attract additional fluxtubes. More generally, configurations containing several defects are known to be frustrated \citep{Drummond2017I,DrummondL2018II,Doran2020,Magistrelli_2026pra}, so many shallow local minima are expected and finding the true, possibly degenerate, ground state is not straightforward. 
Repeating the same numerical experiment with different initial guesses for the PyTorch optimiser gives somewhat different bouquets and surrounding arrangements, but in all the runs considered we find between $6$ and $9$ fluxtubes concentrated around the vortex.

Apart from the detailed structure of all possible bouquets and their associated frustrated configurations, the qualitative point we want to highlight is their effect on the magnetic dressing of a vortex. Vortex-fluxtube pinning, which can be regarded as the simplest bouquet containing a single fluxtube, already magnetizes the core of a vortex in the absence of entrainment. A bouquet can enhance this effect further, producing a magnetic flux around the vortex that can exceed the entrainment-induced magnetic flux of an isolated neutron vortex, which is typically only a fraction of a flux quantum~\citep{Sedrakian1980,Alpar84a,alpar1984ApJ,mendell1991}.

\begin{figure*}[ht]
	\begin{subfigure}{0.32\textwidth}
		\includegraphics[width=\linewidth]{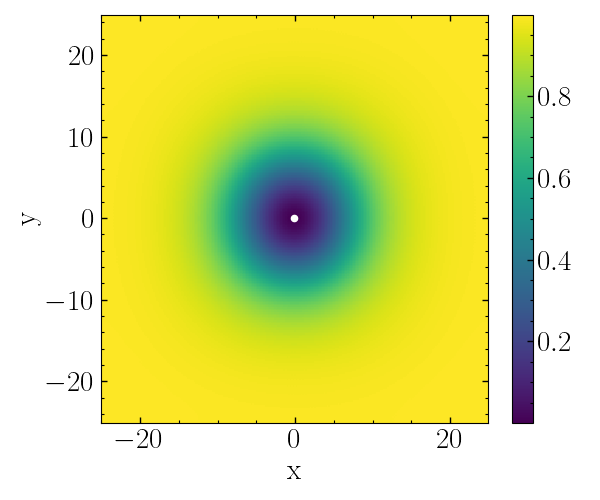}
		\caption{Neutron condensate~$|\psi_n|^2$. }
		\label{fig:psip_type1}
	\end{subfigure}
	\hfill 
	\begin{subfigure}{0.32\textwidth}
		\includegraphics[width=\linewidth]{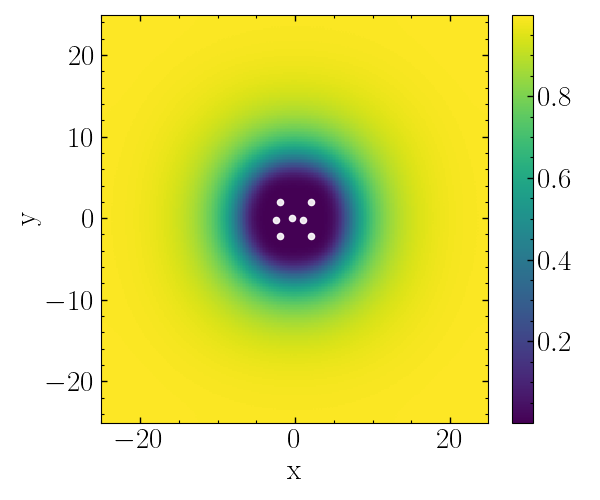}
		\caption{Proton condensate~$|\psi_p|^2$. }
		\label{fig:psin_type1}
	\end{subfigure}
	\hfill 
	\begin{subfigure}{0.32\textwidth}
		\includegraphics[width=\linewidth]{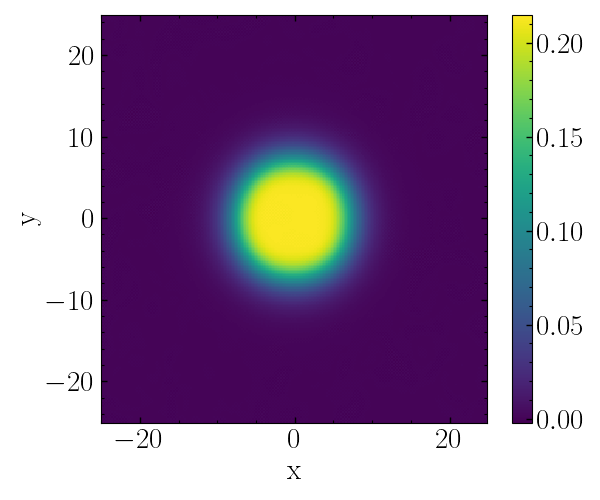}
		\caption{Magnetic field~$B_z$. }
		\label{fig:bz_type1}
	\end{subfigure}
	\caption{Dense bouquet resembling a multiply charged fluxtube ($\mathcal Q_n=1$, $\mathcal Q_p =9$, $\alpha=-0.1$, $g=0$). The white dots represent the positions of the phase singularities within the corresponding condensate. In particular, the 9 phase singularities of $\theta_p$ are all distinct, although it is impossible to distinguish the individual fluxtubes by looking only at the condensate amplitude $|\psi_p|^2$ or~$B_z$.}
	\label{fig:vx_mediated_type_1}
\end{figure*}

In all the simulations performed here, we do not identify a stable multiply charged fluxtube for any combination of the coupling parameters $\alpha$ and $g$ considered. This differs from the multi-quantum fluxtubes found by \citet{alford_good_PRB} in other regions of parameter space. There is, however, a regime in which a bouquet can resemble a multiply charged fluxtube if one considers only the condensate amplitudes and magnetic field: when a very tight bouquet forms, it produces an extended region of proton-condensate depletion in which magnetic flux accumulates. This is illustrated in Fig.~\ref{fig:vx_mediated_type_1}, where a single vortex forms a tight bouquet of 9 fluxtubes for $\alpha=-0.1$. In this case, the coupling is strong enough that the individual fluxtubes are essentially impossible to distinguish from $|\psi_p|^2$ or $B_z$ alone. Their topological identity becomes clear only by inspecting $\theta_p$, which shows nine distinct proton phase singularities.

\subsection{Type-1.5 fluxtube clustering}
\label{sec_type15}

In a single-component type-II superconductor described by standard Ginzburg-Landau modelling, fluxtubes of equal polarization repel and form an Abrikosov lattice in a clean sample. In the coupled system studied here, the fluxtube-fluxtube force can instead become non-monotonic: it is repulsive at short distances and attractive at intermediate distances. We use this non-monotonic interaction as a local diagnostic of type-1.5-like behaviour. 

\begin{figure}
	\centering
	\includegraphics[width=\linewidth]{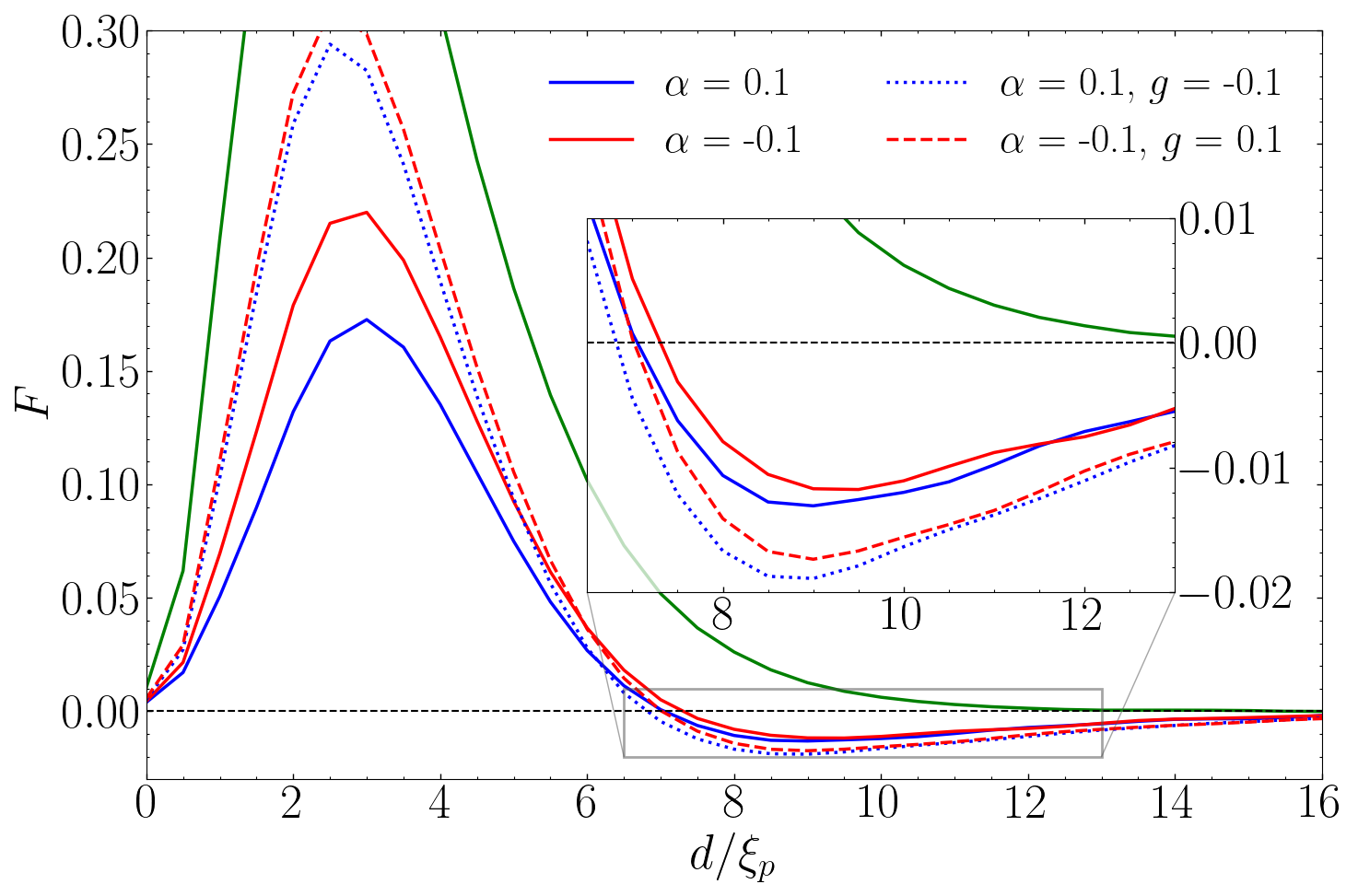}
	\caption{The fluxtube-fluxtube force per unit length $f_{pp}(d)$ in~\eqref{eq_fpp_definition} for representative values of $\alpha$ and $g$. The uncoupled case ($\alpha=g=0$, solid green curve) is purely repulsive over the range shown. For sufficiently large coupling, $f_{pp}(d)$ changes sign at finite separation, defining the distance $d_\star$ used in the comparison with the many-fluxtube clusters in Fig.~\ref{fig_type15_clusters}.
	}
	\label{fig_type15_force}
\end{figure}

Fig.~\ref{fig_type15_force} shows the effective force between two fluxtubes for representative values of $\alpha$ and $g$, extracted using the same fixed-separation strategy as in Fig.~\ref{fig_interaction_schematic} but imprinting the phase of two fluxtubes at distance $d$. Unlike the vortex-fluxtube case, we do not subtract the uncoupled configuration, since two fluxtubes also interact for $\alpha=g=0$. Denoting by $F_{pp}(d;\alpha,g)$ the constrained minimum with two fluxtubes at separation $d$, we define
\begin{equation}
	E_{pp}(d;\alpha,g)
	=
	F_{pp}(d;\alpha,g)
	-
	F_{pp}(d_{\rm ref};\alpha,g),
	\label{eq_Epp_definition}
\end{equation}
and the corresponding force per unit length
\begin{equation}
	f_{pp}(d)
	=
	-E'_{pp}(d).
	\label{eq_fpp_definition}
\end{equation}
For sufficiently large coupling the effective force changes sign at a finite separation. We denote by $d_\star$ the corresponding preferred inter-fluxtube distance, namely the separation at which the two-fluxtube $E_{pp}(d;\alpha,g)$ has a local minimum; i.e., $f_{pp}(d_\star)=0$.

There is no contradiction between the results in Fig.~\ref{fig_type15_force} and the local interpretation of $\alpha>0$ as repulsive in Sec.~\ref{sec_rep_attr}: here the attraction is an emergent non-local interaction between two proton fluxtubes, mediated by the coupled response of both condensates. Furthermore, the fluxtubes always remain repulsive at short separations within the explored range of $\alpha$ and~$g$.

\begin{figure*}
	\centering
	\begin{subfigure}{0.48\textwidth}
		\centering
		\includegraphics[width=\linewidth]{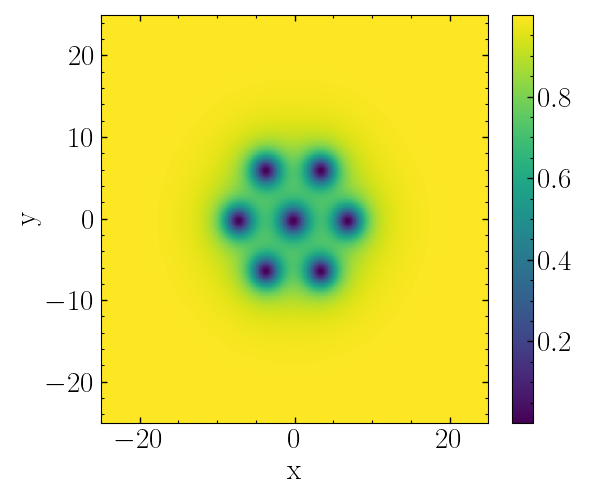}
		\caption{Fluxtube cluster with $\mathcal Q_n=0$, $\mathcal Q_p =7$.}
		\label{fig_type15_np_7}
	\end{subfigure}
	\hfill
	\begin{subfigure}{0.48\textwidth}
		\centering
		\includegraphics[width=\linewidth]{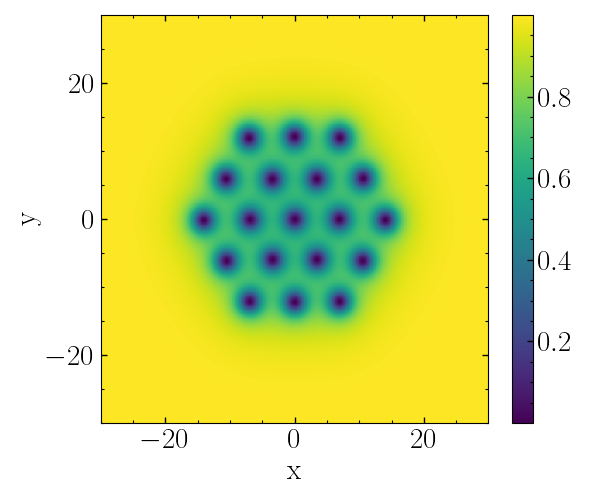}
		\caption{Fluxtube cluster with $\mathcal Q_n=0$, $\mathcal Q_p =19$.}
		\label{fig_type15_np_19}
	\end{subfigure}
	\caption{Two examples of type-1.5 fluxtube clusters at $\alpha=0.1$ and $g=0$, obtained with QPBC and $\mathcal Q_n=0$, $\mathcal Q_p =7,19$. Both panels show the proton condensate amplitude $|\psi_p|^2$, with full condensate depletion at the centre of each distinct fluxtube core. The fluxtube positions are not fixed and no hexagonal initial configuration is imposed. The clusters self-assemble from a random initial condition, and the resulting nearest-neighbour separation is consistent with the preferred distance inferred from the two-fluxtube force in Fig.~\ref{fig_type15_force}.}
	\label{fig_type15_clusters}
\end{figure*}

We then ask whether this two-fluxtube diagnostic survives in a genuine unconstrained many-fluxtube calculation. Essentially, we want to see whether many fluxtubes that are free to arrange themselves within $\mathcal D$ do so with a characteristic spacing close to $d_\star$.
To this end, we minimize configurations with several fluxtubes using QPBC, fixing only the total fluxtube number $\mathcal{Q}_p$ and not their positions (unlike in the bouquet case, we now have $\mathcal Q_n=0$). For $\alpha=0.1$, the fluxtubes do not disperse uniformly over the computational domain. Instead, they form compact clusters while remaining individually resolved, as shown in Fig.~\ref{fig_type15_clusters}. No initial configuration with hexagonal geometry is imposed: the fluxtubes self-arrange during the minimisation from a random initial condition.

The interesting point is the comparison between Figs.~\ref{fig_type15_force} and~\ref{fig_type15_clusters}. The typical nearest-neighbour separation inside the QPBC clusters is consistent with the preferred separation $d_\star$ inferred independently from the two-fluxtube force curve. 
Visually, the $\mathcal Q_p=7$ cluster has a nearest-neighbour spacing $d_{\rm nn}\simeq7.5$, corresponding to about $235\,{\rm fm}$, while the $\mathcal Q_p=19$ cluster gives $d_{\rm nn}\simeq7.0-7.2$, or about $220-226\,{\rm fm}$. These values are close to the first zero of the two-fluxtube force in Fig.~\ref{fig_type15_force}, $d_\star\simeq7$, corresponding to about $220\,{\rm fm}$, but are not identical. The two-fluxtube calculation therefore identifies a characteristic clustering scale $d_\star$ rather than the exact lattice constant.

Clustered type-1.5-like fluxtube phases in a superconductor coupled to a superfluid were already discussed by \citet{haber2017prd} and demonstrated numerically with QPBC by \citet{Wood_2022Univ}, whose calculations include genuine entrainment. Our calculation with QPBC shows that the same type-1.5 mechanism persists in the reduced non-entrained functional used here, and that the preferred separation $d_\star$ from the two-fluxtube calculation provides a useful estimate of the spacing $d_{\rm nn}$ in self-assembled many-fluxtube clusters. 

\section{Discussion}

The possibility that a neutron vortex interacts with, or is pinned to, one or more proton fluxtubes has been considered extensively \citep{Sedrakyan1991JETP,Srinivasan2,Ruderman1998,alpar2017JApA,sourie2020_pinningCore,souriechamel2020,Drummond2017I,DrummondL2018II,melatos_III_2023}. Three-dimensional simulations also show simultaneous pinning of different vortex segments to different fluxtubes \citep{Drummond2017I,DrummondL2018II,melatos_III_2023}, and more recent simulations evolve both condensates with attractive vortex-fluxtube interactions \citep{Shukla_PRD_2024}. The bouquet structure found here is more specific: a finite number of pre-existing fluxtubes collect around a neutron vortex because vortex-fluxtube attraction competes with fluxtube-fluxtube repulsion. To our knowledge, such a small vortex-centred bound aggregate has not been identified explicitly in previous studies. It is also distinct from the type-1.5-like clusters of Sec.~\ref{sec_type15}, which do not require a neutron vortex~\citep{haber2017prd,Wood_2022Univ}.

At this point it is useful to distinguish the small ``bouquets'' found here from the ``vortex clusters'' introduced by \citet{Sedrakian1995ApJ}. Both structures involve a neutron vortex surrounded by proton fluxtubes, but their origin and characteristic scales are very different. In the vortex-cluster model of \citet{Sedrakian1995ApJ}, genuine neutron-proton entrainment (absent in our model) makes the circulation around a neutron vortex induce a proton superconducting current and an associated magnetic field: this is a long-distance effect that can magnetize matter in a relatively large area around the vortex. In this region, the entrainment-induced field may overcome the critical field for fluxtube nucleation, so the proton superconductor locally enters the mixed state. The resulting vortex cluster is therefore a comparatively large mixed-state region surrounding the neutron vortex; for their representative microscopic parameters, \citet{Sedrakian1995ApJ} find a cluster radius of order $10^{-5}\,\mathrm{cm}$ and as many as $\sim 10^{12}$ fluxtubes associated with a single neutron vortex.

The bouquets considered here are instead small composite defects with a different origin, clustered within a radius of $\sim 10\xi_p$ or less, depending on the size of the bouquet. Genuine phase-gradient entrainment is absent from \eqref{eq_dimensionless_free_energy_density}; the proton fluxtubes are already present because the magnetic flux is imposed through the QPBC, and their attraction to the neutron vortex is generated by the local density-density and density-gradient couplings $\alpha$ and $g$. The competition between this short-range vortex-fluxtube attraction and fluxtube-fluxtube repulsion produces a small number of nearby bound fluxtubes, of order unity in the configurations studied here. Thus, a bouquet is not the small-scale limit of the vortex cluster of~\citet{Sedrakian1995ApJ}.

We now turn to the possible relevance for neutron star evolution. We first ask whether the dragging of bouquets by vortices may have an observable imprint during glitches. A simple order-of-magnitude estimate illustrates the scale of the magnetic-field change that would occur if vortices dragged their associated bouquet flux during a glitch with angular velocity jump $\Delta \Omega$. 
If, on average, $P$ fluxtubes are transported per vortex, then\footnote{
The estimate assumes co-motion of the bound fluxtubes with vortices and a spatially uniform magnetic-field response in planar geometry; we use it only to illustrate the overall scaling of the transported flux.}
\begin{equation}
	\Delta B \sim 2P\frac{\Delta\Omega}{\kappa_n}\Phi_0
	\sim
	10^{-8}\,
	P
	\left(
	\frac{\Delta\Omega}{ 10^{-4}\,{\rm rad\,s^{-1}}}
	\right)
	{\rm G}
	\label{eq_delta_b}
\end{equation}
where $\kappa_n =  2\pi\hbar/(2 m_n) \simeq 1.99\times10^{-3}\,{\rm cm^2\,s^{-1}}$ is the neutron circulation quantum and $\Phi_0=  2\pi\hbar c/(2e) \simeq2.07\times10^{-7}\,{\rm G\,cm^2}$ is the proton flux quantum.
For $\Delta\Omega$ slightly larger than that of the largest Vela glitch and for average bouquets with $P=9$, this gives only $\Delta B \sim 10^{-6}\,{\rm G}$: the direct effect of bouquet transport on the global magnetic field during a glitch seems unobservable, unless the dragging is concentrated in a small region (here we assumed a homogeneous readjustment of the vortex configuration).

The more relevant consequence of a bouquet may thus be local dissipation rather than a measurable effect on the magnetic-field configuration. A bouquet magnetizes the vicinity of a vortex and can therefore alter electron scattering around its core. This may modify the vortex-mediated hydrodynamic mutual friction \citep{alpar1984ApJ,AndSid06}. In particular, the calculation by \citet{sauls_1982} of the coupling timescale between the superfluid and the charged component due to electron scattering could be revisited using the magnetic structure of a bouquet, if such structures also survive in the parameter space of a microscopically calibrated Ginzburg-Landau functional.

Another potentially interesting point is the connection with rotochemical heating. In the model of \citet{rodriguez_reisenegger_2026}, proton superconductivity suppresses Urca reactions except in magnetized regions where the proton condensate is absent, such as fluxtube cores in type-II matter or normal domains in type-I matter.  
Larger-scale QPBC calculations, once microscopically calibrated, could provide mesoscopic configurations of the proton condensate and magnetic field that determine how the reaction-active volume is distributed. In particular, vortex-centred bouquets and type-1.5 fluxtube clusters can concentrate proton-depleted, magnetized regions locally. A quantitative application to rotochemical heating would require computing an effective local suppression factor for the Urca rates from the profiles of $|\psi_p|^2$ and $B_z$, and then incorporating it into a stellar thermal-evolution model.
 
Finally, we stress that our two-dimensional results are local and do not imply globally straight vortices or fluxtubes. 
Neither of these two families of topological defects is expected to arrange in an ordered array at the stellar scale, although this is a common, often implicit, working assumption of many glitch models as discussed in \citep{AMP_arxiv_2023}: three-dimensional simulations have shown that vortices and fluxtubes can form complicated locally pinned structures and tangles \citep{Drummond2017I,DrummondL2018II,melatos_III_2023}. 
Thus, the present calculations apply only where a local patch can be approximated by nearly parallel line defects, so that a local separation and interaction energy per unit length are meaningful. Misaligned defects require a three-dimensional treatment, for which QPBC can also be formulated~\citep{Wood2019PhRvB}.

\section{Conclusions}
\label{sec_conclusions}

We have considered a phenomenological Ginzburg-Landau free energy functional for two interacting scalar condensates at zero temperature. The model is closely related to that of \citet{alford_good_PRB} and corresponds to a reduced no-entrainment sector of the more general models of \citep{alpar1984ApJ}, \citet{haber2017prd} and \citet{Wood_2022Univ}; see App.~\ref{app_model_dictionary}. 

We performed two kinds of numerical experiments with different constraints. First, we used QPBC to impose the net numbers of neutron vortices and proton fluxtubes while allowing the condensate fields to relax. This gives the single-defect profiles and the binding-energy proxy of Fig.~\ref{fig_klausner_proxy}. We also used constrained phase imprinting to determine the separation-dependent vortex-fluxtube interaction. Attractive couplings, in the local sense of Sec.~\ref{sec_rep_attr}, favour overlapping vortex-fluxtube states and produce local pinning.

A further application of QPBC is the exploration of configurations in which one neutron vortex coexists with several fluxtubes. We find that competition between vortex-fluxtube attraction and fluxtube-fluxtube repulsion can produce fluxtube bouquets around the vortex. These structures are different from the much larger entrainment-induced vortex clusters of \citet{Sedrakian1995ApJ}. Even in the densest bouquets studied here, the proton phase shows distinct fluxtube singularities, so the configuration remains a collection of singly quantised fluxtubes rather than a multiply charged defect.

Even in the absence of a neutron vortex, sufficiently strong couplings within our working parameter range can produce short-range fluxtube repulsion and intermediate-range attraction, leading to type-1.5-like clustered configurations \citep{haber2017prd,Wood_2022Univ}. The characteristic nearest-neighbour spacing in the relaxed clusters is consistent with the preferred distance $d_\star$ that we obtained from the two-fluxtube interaction.

Bouquet and type-1.5-like configurations may be relevant for rotochemical heating \citep{rodriguez_reisenegger_2026} and dissipative mutual friction in the outer core \citep{sauls_1982,alpar1984ApJ,AndSid06}, because they change the local distribution of proton-condensate depletion and magnetic flux. These implications remain qualitative in the present study, since we do not calculate Urca rates or electron transport coefficients in realistic neutron star matter.

Finally, the coupling parameters $\alpha$ and $g$ remain poorly constrained in neutron star matter. Predictive applications require microscopic input for these couplings, the inclusion of genuine phase-gradient entrainment, and a three-dimensional treatment of curved or misaligned defects, as in \citet{melatos_III_2023}.

\begin{acknowledgements}
	We thank Michael Urban and Vitali Ungurean for useful discussions, and Sanchana Krishna for the diagrams in Figs.~\ref{fig_interaction_schematic} and~\ref{fig_klausner_proxy}. 
	We acknowledge the support of the IN2P3 Master Project NewMAC and MAC, the ANR project `Gravitational waves from hot neutron stars and properties of ultra-dense matter' (GW-HNS, ANR-22-CE31-0001-01), and the CNRS International Research Project (IRP) `Origine des \'el\'ements lourds dans l'univers: Astres Compacts et Nucl\'eosynth\`ese (ACNu)'.
	Partial support comes from the Cost Action CA24139 - Superfluid Condensates in Astrophysics and Laboratory Experiments (SCALES).
\end{acknowledgements}

\appendix

\section{Rescaling and dimensionless variables}
\label{app_rescaling}

The dimensionless energy cost of a stationary configuration follows from the rescaling procedure in \citep{alford_good_PRB,Wood_2022Univ}. We denote the dimensional condensate fields by $\Psi_x(\bm r)$, normalised such that the squared modulus $|\Psi_x|^2$ has dimensions of a number density. The dimensional energy density of the condensate sector is~\citep{alford_good_PRB,haber2017prd,Wood_2022Univ}
\begin{align}
\tilde{\mathcal{F}} =& \sum_{x=n,p}
\left[
\frac{\hbar^2}{2m_x^*}\left|D_x\Psi_x\right|^2
+ \frac{\lambda_x}{2}\left(|\Psi_x|^2-n_x^*\right)^2
\right]
\notag\\ & +
\tilde{\alpha}
\left(|\Psi_n|^2-n_n^*\right)
\left(|\Psi_p|^2-n_p^*\right)
\notag\\ & +
\frac{\tilde{g}\hbar^2}{4m_n^*m_p^*}
\nabla|\Psi_p|^2\cdot\nabla|\Psi_n|^2 
\notag\\ & +
\frac{1}{8\pi}\left|\nabla\times \tilde{\bm{A}}\right|^2 ,
\label{eq_dimensional_free_energy_compact}
\end{align}
where $m_x^*$ are the condensate masses, and
\begin{equation}
	D_x
	=
	\nabla
	-
	i \frac{q_x^*}{\hbar c}\tilde{\bm{A}},
	\qquad
	q_n^*=0,
	\qquad
	q_p^*=2e.
\end{equation}
The presence of non-zero couplings $\tilde{\alpha}$ and $\tilde{g}$ does not modify the form of the usual single-condensate $U(1)_x$ currents.
The usual single-condensate\footnote{
	Namely, considering the uncoupled limit of \eqref{eq_dimensional_free_energy_compact}. For the coupled theory they remain useful typical length scales but lose their original meaning, and similarly for $\kappa$ in~\eqref{eq_fagiano_cruento}.} 
coherence length $\xi_x$ (the single-condensate healing length is $\xi_x/\sqrt{2}$) and magnetic penetration length $\ell_p$ are defined according to~\citep{tinkham2004introduction} as
\begin{equation}
	\label{eq_penetration}
	\xi_x^2 =\frac{\hbar^2} { 2 m_x^*\lambda_x n^*_x }
	,\qquad
	\ell_p^2  =  \frac{m_p^*\, c^2}
	{4\pi \, q_p^{*2} \, n^*_p},
\end{equation}
while the fields can be rescaled according to
\begin{equation}
	\Psi_x=\sqrt{n_x^*}\,\psi_x,
	\qquad
	\tilde{\bm A}
	=
	\frac{\hbar c}{q_p^*\xi_p}\bm{A},
\end{equation}
and, following \citep{alford_good_PRB,Wood_2022Univ}, we use dimensionless spatial coordinates $\bm x$ and a dimensionless free energy density $\mathcal{F}$ defined as 
\begin{equation}
	\bm{r}=\xi_p\bm{x},
	\qquad
	\tilde{\mathcal{F}}
	= \lambda_p n_p^{*2}\,\mathcal{F} .
\end{equation}
It is convenient to introduce the dimensionless parameters~\citep{Wood_2022Univ}
\begin{equation}
	\label{eq_fagiano_cruento}
	\epsilon = \frac{m_n^*n_p^*}{m_p^*n_n^*},
	\qquad
	R = \frac{\xi_p}{\xi_n},
	\qquad
	\kappa = \frac{\ell_p}{\xi_p},
\end{equation}
and
\begin{equation}
	\alpha
	=
	\frac{m_n^*}{m_p^*}\frac{\tilde{\alpha}}{\lambda_p},
	\qquad
	g
	=
	\frac{\tilde{g}n_p^*}{2m_p^*}.
\end{equation}
With this rescaling, one obtains exactly the $\mathcal{F}$ in \eqref{eq_dimensionless_free_energy_density}, where $\bm \nabla$ indicates derivatives in $\bm x$.
For configurations invariant along the $z$-direction, the dimensional energy per unit length is
\begin{equation}
	\tilde{F} = \lambda_p n_p^{*2}\xi_p^2 F = \lambda_p n_p^{*2}\xi_p^2
	\int_{\mathcal D}\mathcal{F} \,dx dy \, ,
\end{equation}
where the dimensionless free energy per unit length $F$ is given in~\eqref{eq_free_energy}.
Similarly, the dimensional pinning energy per unit length $\tilde E_{\rm pin}$ is obtained as
\begin{equation}
\label{eq_fagianocongilione}
	\tilde E_{\rm pin} 
	=
	\lambda_p n_p^{*2}\xi_p^2 E_{\rm pin}
	= \frac{(\hbar c)^2 n_p}{4m_p^* c^2}\,E_{\rm pin},
\end{equation}
where $n_p^*=n_p/2$. 
For the fiducial conversion used here, we adopt the benchmark values in \citep{Wood_2022Univ}: we take $m_p^*=2m$, with $m c^2 = 931\,{\rm MeV}$, and $n_p=0.0251\,{\rm fm}^{-3}$. 
With this choice, \eqref{eq_fagianocongilione} gives
\begin{equation}
	\label{eq_facocero}
	\tilde E_{\rm pin}
	\simeq
	0.131\,E_{\rm pin}\,\,{\rm MeV\,fm^{-1}} .
\end{equation}

\section{Dictionary between stationary two-condensate models for neutron star cores}
\label{app_model_dictionary}

We compare several two-condensate models in the stationary limit, after removing genuine phase-gradient entrainment, which is the case considered in \eqref{eq_dimensionless_free_energy_density} or, equivalently, \eqref{eq_dimensional_free_energy_compact}. 
In this section, all models are dimensional and $\mathcal{F}$ has physical dimensions of an energy density; the tilde notation of Sec.~\ref{app_rescaling} is dropped to ease the notation. 
We use $x,y=n,p$ as species indices (repeated indices are summed) and define
\begin{equation}
	N_x = |\Psi_x|^2,
	\qquad
	\delta N_x = N_x-N_x^{0},
\end{equation}
where $N_x^{0}$ is the homogeneous equilibrium value of the corresponding order parameter. Neglecting phase-gradient entrainment, a convenient and general form for the free energy density is
\begin{multline}
	\mathcal F
	=
	\frac{1}{8\pi}
	|\nabla\times\bm A|^2
	+
	\mathsf K_{xy}
	(D_x\Psi_x)^*\cdot D_y\Psi_y
	\\+
	\frac{1}{2}
	\mathsf\Lambda_{xy}
	\delta N_x\delta N_y
	+
	\frac{1}{2}
	\mathsf C_{xy}
	\nabla N_x\cdot\nabla N_y ,
	\label{eq_canonical_stationary_functional}
\end{multline}
where repeated species indices are summed, and $\mathsf K$,  $\mathsf\Lambda$ and $\mathsf C$ are real symmetric matrices. Moreover, \eqref{eq_canonical_stationary_functional} is invariant under $U(1)_n \times U(1)_p$ if $\mathsf K$ is diagonal.
The covariant derivatives are
\begin{equation}
	D_x
	=
	\nabla
	-
	i\frac{q_x}{\hbar c}\bm A,
\end{equation}
with $q_x$ and the field normalisations left unspecified: the only dimensional constraint is that $\mathcal F$ has physical dimensions of an energy density.
Different additive constants that may be present in the free energy models below are omitted throughout. To build a simple dictionary between them, they are all mapped into~\eqref{eq_canonical_stationary_functional}.
\\
\\
\emph{Present model} -- Dropping the tilde in \eqref{eq_dimensional_free_energy_compact} to ease the notation, our dimensional model is
\begin{multline}
	\label{eq_cicciopazzo}
	\mathcal F
	=  \sum_{x=n,p} \mathcal E_x +
	\alpha
	\left(|\Psi_n|^2-n_n^*\right)
	\left(|\Psi_p|^2-n_p^*\right)
	\\
	+
	\frac{g\hbar^2}{4m_n^*m_p^*}
	\nabla|\Psi_n|^2\cdot
	\nabla|\Psi_p|^2 + \frac{1}{8\pi} |\nabla\times\bm A|^2.
\end{multline}
Here $\mathcal E_x$ are the single-condensate terms defined by the first row in \eqref{eq_dimensional_free_energy_compact}.
The free energy density \eqref{eq_canonical_stationary_functional} is therefore obtained with $N_x^{0}=n_x^*$ and
\begin{equation}
	\label{eq_struttura}
	\mathsf K
	=
	\begin{pmatrix}
		\dfrac{\hbar^2}{2m_n^*} & 0                       
		\\[2mm]
		0                       & \dfrac{\hbar^2}{2m_p^*} 
	\end{pmatrix},
	\qquad
	\mathsf\Lambda
	=
	\begin{pmatrix}
		\lambda_n & \alpha    
		\\
		\alpha    & \lambda_p 
	\end{pmatrix},
\end{equation}
\begin{equation}
	\mathsf C
	=
	\begin{pmatrix}
		0 &   
		\dfrac{g\hbar^2}{4m_n^*m_p^*}
		\\[2mm]
		\dfrac{g\hbar^2}{4m_n^*m_p^*}
		  &   
		0
	\end{pmatrix}.
\end{equation}
\\
\\
\emph{Relation with \citep{Wood_2022Univ} } -- \citet{Wood_2022Univ} introduce four derivative couplings $h_1,\ldots,h_4$. Their phase-gradient entrainment is controlled by $h_1$, and the no-entrainment limit is obtained by setting $h_1=0$.
Before imposing their additional simplifications $m_n=m_p=m $ and $h_3=h_4$, their stationary free energy density takes the form
\begin{align}
	\label{eq_woodgraber}
	\mathcal F  
	={} &   
	\frac{g_{pp}}{2}
	\left(|\psi_p|^2-\frac{n_p}{2}\right)^2
	+
	\frac{g_{nn}}{2}
	\left(|\psi_n|^2-\frac{n_n}{2}\right)^2
	\notag\\
	    & + 
	g_{pn}
	\left(|\psi_p|^2-\frac{n_p}{2}\right)
	\left(|\psi_n|^2-\frac{n_n}{2}\right)
	+
	\frac{|\nabla\times\bm A|^2}{8\pi}
	\notag\\
	    & + 
	\frac{\hbar^2}{4m_p}|D_p\psi_p|^2
	+
	\frac{\hbar^2}{4m_n}|\nabla\psi_n|^2
	+
	\frac{h_2}{2}
	\nabla|\psi_p|^2\cdot\nabla|\psi_n|^2
	\notag\\
	    & + 
	\frac{h_3}{4}
	\left|\nabla|\psi_p|^2\right|^2
	+
	\frac{h_4}{4}
	\left|\nabla|\psi_n|^2\right|^2 .
\end{align}
Hence, the mapping with~\eqref{eq_canonical_stationary_functional} is obtained as
\begin{equation}
	\Psi_x=\psi_x,
	\qquad
	N_n^{0}=n_n/2,
	\qquad
	N_p^{0}= n_p/2,
\end{equation}
where $n_n$ and $n_p$ are the actual number densities of neutrons and protons. The mapping is given by
\begin{equation}
	\mathsf K  
	=
	\begin{pmatrix}
		\dfrac{\hbar^2}{4m_n} & 0                     
		\\[2mm]
		0                     & \dfrac{\hbar^2}{4m_p} 
	\end{pmatrix},
	\, \,
	\mathsf\Lambda  
	=
	\begin{pmatrix}
		g_{nn} & g_{pn} 
		\\
		g_{pn} & g_{pp} 
	\end{pmatrix},
	\, \, 
	\mathsf C  
	=
	\frac{1}{2}
	\begin{pmatrix}
		h_4 & h_2 
		\\
		h_2 & h_3 
	\end{pmatrix}.
\end{equation}
\citet{Wood_2022Univ} subsequently impose $m_n=m_p=m $ and $h_3=h_4$, but their no-entrainment limit is therefore more general than the present one because it retains the diagonal terms in~$\mathsf C$.
\\
\\
\emph{Relation with \citep{haber2017prd}} -- \citet{haber2017prd} start from a relativistic model with $\hbar=c=1$, and denote the charged and neutral fields by $\varphi_1$ and $\varphi_2$, respectively. 
In their static free energy, $g$ multiplies the mixed phase-gradient term, whereas $G$ multiplies the mixed density-gradient term. Genuine phase entrainment is therefore removed by imposing $g=0$.
At fixed temperature, their stationary free energy density then becomes
\begin{align}
	\mathcal F 
	={} &   
	|D_1\varphi_1|^2
	+
	|\nabla\varphi_2|^2
	+
	\frac{|\nabla\times\bm A|^2}{8\pi}
	\notag\\
	    & + 
	(m_{1}^2-\mu_1^2)|\varphi_1|^2
	+
	(m_{2}^2-\mu_2^2)|\varphi_2|^2
	\notag\\
	    & + 
	\lambda_1|\varphi_1|^4
	+
	\lambda_2|\varphi_2|^4
	-
	2h |\varphi_1|^2|\varphi_2|^2
	\notag\\
	    & - 
	\frac{G}{2}
	\nabla|\varphi_1|^2\cdot
	\nabla|\varphi_2|^2 .
\end{align}
With the identification $\Psi_p=\varphi_1$ and $\Psi_n=\varphi_2$, the homogeneous coexistence state satisfies
\begin{equation}
	\begin{split}
		2\lambda_1 N_p^{0}
		-
		2h N_n^{0}
		&=
		\mu_1^2-m_1^2,
		\\
		2\lambda_2 N_n^{0}
		-
		2h N_p^{0}
		&=
		\mu_2^2-m_2^2.
	\end{split}
\end{equation}
The mapping into \eqref{eq_canonical_stationary_functional} is therefore completed by the solution to the above system and
\begin{equation}
	\mathsf K 
	=
	\begin{pmatrix}
		1 & 0 \\
		0 & 1 
	\end{pmatrix},
	\,\,
	\mathsf\Lambda 
	=
	\begin{pmatrix}
		2\lambda_2 & -2h        
		\\
		-2h        & 2\lambda_1 
	\end{pmatrix},
	\,\,
	\mathsf C 
	=
	\begin{pmatrix}
		0 & k 
		\\
		k & 0 
	\end{pmatrix},
\end{equation}
with $k=-G/2$. The structure of the above matrices is identical to our~\eqref{eq_struttura}.
\\
\\
\emph{ Relation with \citep{alpar1984ApJ}} -- \citet{alpar1984ApJ} consider the free energy density $\mathcal{F}=f_u+f_g$ with 
\begin{equation}
	f_u
	=
	\alpha_p|\psi_p|^2
	+
	\beta_p|\psi_p|^4
	+
	\alpha_n|\psi_n|^2
	+
	\beta_n|\psi_n|^4
	+
	\nu|\psi_p|^2|\psi_n|^2 ,
\end{equation}
and
\begin{align}
	f_g
	={} &   
	\gamma_p|\nabla\psi_p|^2
	+
	\gamma_n|\nabla\psi_n|^2
	\notag\\
	    & + 
	\mu_1
	(\nabla\psi_p\cdot\nabla\psi_n^*)
	\psi_p^*\psi_n
	+
	\mu_2
	(\nabla\psi_p^*\cdot\nabla\psi_n)
	\psi_p\psi_n^*
	\notag\\
	    & + 
	\mu_3
	(\nabla\psi_p\cdot\nabla\psi_n)
	\psi_p^*\psi_n^*
	+
	\mu_4
	(\nabla\psi_p^*\cdot\nabla\psi_n^*)
	\psi_p\psi_n ,
\end{align}
where $ \mu_1=\mu_2^*$ and $\mu_3=\mu_4^*$.
The time-reversal-even sector corresponds to real $\mu_i$, so that $\mu_1=\mu_2$, $\mu_3=\mu_4$.
After restoring minimal coupling for the protons and writing $ \psi_x=\rho_x \exp{(i\theta_x)}$, the mixed-gradient contribution becomes
\begin{align}
	f_{g,\mathrm{mix}}
	={} &   
	2(\mu_1+\mu_3)
	\rho_p\rho_n
	\nabla\rho_p\cdot\nabla\rho_n
	\notag\\
	    & + 
	2(\mu_1-\mu_3)
	\rho_p^2\rho_n^2
	\bm P_p\cdot\bm P_n ,
\end{align}
where
\begin{equation}
	\bm P_p
	=
	\nabla\theta_p
	-
	\frac{2e}{\hbar c}\bm A,
	\qquad
	\bm P_n
	=
	\nabla\theta_n.
\end{equation}
The second term is the genuine phase-gradient entrainment. Removing it therefore requires
$ \mu_1=\mu_3$. Together with the reality conditions this gives
\begin{equation}
	\mu_1=\mu_2=\mu_3=\mu_4\equiv\mu.
\end{equation}
The remaining mixed-gradient term then reduces exactly to
\begin{equation}
	f_{g,\mathrm{mix}}
	=
	\mu\,
	\nabla|\psi_p|^2\cdot
	\nabla|\psi_n|^2.
\end{equation}
Let $N_x^{0}$ denote the homogeneous coexistence value of $|\psi_{x}|^2$. The equilibrium conditions are
\begin{equation}
	\alpha_p
	+
	2\beta_p N_p^{0}
	+
	\nu N_n^{0}
	=0,
\end{equation}
\begin{equation}
	\alpha_n
	+
	2\beta_nN_n^{0}
	+
	\nu N_p^{0}
	=0.
\end{equation}
Up to an additive constant, $f_u$ can therefore be written in the shifted form of \eqref{eq_canonical_stationary_functional}. The dictionary is
\begin{equation}
	\mathsf K 
	=
	\begin{pmatrix}
		\gamma_n & 0        
		\\
		0        & \gamma_p 
	\end{pmatrix},
	\, \, 
	\mathsf\Lambda 
	=
	\begin{pmatrix}
		2\beta_n & \nu      
		\\
		\nu      & 2\beta_p 
	\end{pmatrix}, \, \, 
	\mathsf C 
	=
	\begin{pmatrix}
		0   & \mu 
		\\
		\mu & 0   
	\end{pmatrix}.
\end{equation}
Again, the structure of the above matrices is identical to~\eqref{eq_struttura}.
\\
\\
\emph{Relation with \citep{alford_good_PRB}} -- 
The dimensional free energy density of \citet{alford_good_PRB} is
\begin{align}
	\mathcal F
	={} &   
	\frac{\hbar^2}{2m_c}
	\left[
	\left|
	\left(
	\nabla-\frac{iq}{\hbar c}\bm A
	\right)\phi_p
	\right|^2
	+
	|\nabla\phi_n|^2
	\right]
	+
	\frac{|\nabla\times\bm A|^2}{8\pi}
	\notag\\
	    & + 
	\frac{a_{pp}}{2}
	\left(
	|\phi_p|^2-\langle\phi_p\rangle^2
	\right)^2
	+
	\frac{a_{nn}}{2}
	\left(
	|\phi_n|^2-\langle\phi_n\rangle^2
	\right)^2
	\notag\\
	    & + 
	a_{pn}
	\left(
	|\phi_p|^2-\langle\phi_p\rangle^2
	\right)
	\left(
	|\phi_n|^2-\langle\phi_n\rangle^2
	\right)
	+
	U_{\rm ent} ,
\end{align}
with $m_c=2m$ and $  q=2e$.
Although $U_{\rm ent} $ is referred to as an entrainment interaction, their expression reduces to
\begin{equation}
	U_{\rm ent} 
	=
	-
	\frac{\hbar^2\sigma}
	{4m_c\langle\phi_p\rangle\langle\phi_n\rangle}
	\nabla|\phi_p|^2\cdot
	\nabla|\phi_n|^2.
\end{equation}
It therefore contains no genuine phase-gradient entrainment and already belongs to the class described by \eqref{eq_canonical_stationary_functional}. The mapping is
\begin{equation}
	\Psi_x=\phi_x,
	\qquad
	N_x^{0}=\langle\phi_x\rangle^2,
\end{equation}
with
\begin{equation}
	\mathsf K 
	=
	\frac{\hbar^2}{2m_c}
	\begin{pmatrix}
		1 & 0 \\
		0 & 1 
	\end{pmatrix},
	\qquad
	\mathsf\Lambda 
	=
	\begin{pmatrix}
		a_{nn} & a_{pn} 
		\\
		a_{pn} & a_{pp} 
	\end{pmatrix},
\end{equation}
and
\begin{equation}
	\mathsf C 
	=
	\begin{pmatrix}
		0 & C 
		\\
		C & 0 
	\end{pmatrix},
	\qquad
	C 
	=
	-
	\frac{\hbar^2\sigma}
	{4m_c\langle\phi_p\rangle\langle\phi_n\rangle}.
\end{equation}
The structure of the above matrices is identical to our~\eqref{eq_struttura}, without the need to remove any sector.
\\
\\
\emph{Relation with \citep{melatos_III_2023}} -- The stationary limit of the bulk free energy of \citet{melatos_III_2023} is\footnote{
	In \citep{melatos_III_2023}, the neutron quadratic term is written as $-\mu_n|\psi|^2/m$, with $\mu_n$ identified there as a chemical potential and $|\psi|^2$ as a number density. Thus, we interpret it as a typo and use~$-\mu_n|\psi|^2$.
}
\begin{align}
	\mathcal F 
	={} &   
	\frac{\hbar^2}{2m^*}|\nabla\psi|^2
	+
	\frac{\hbar^2}{2m^*}
	\left|
	\left(
	\nabla
	-
	i\frac{q}{\hbar c}\bm A
	\right)\phi
	\right|^2
	+
	\frac{|\nabla\times\bm A|^2}{8\pi}
	\notag\\
	    & - 
	\mu_n|\psi|^2
	+
	\frac{U_0}{2}|\psi|^4
	+
	\alpha|\phi|^2
	+
	\frac{\beta}{2}|\phi|^4
	+
	\varsigma|\psi|^2|\phi|^2 .
	\label{eq_melatosF}
\end{align}
Here $m^*$ is a Cooper-pair mass and $q=2e$.
Equation \eqref{eq_melatosF} corresponds to the non-entrained limit considered in their proton-feedback simulations~\citep{melatos_III_2023}.
To compare with~\eqref{eq_canonical_stationary_functional}, we identify $\Psi_n=\psi$ and $\Psi_p=\phi$. The homogeneous equilibrium amplitudes satisfy
\begin{equation}
	U_0N_n^{0}
	+
	\varsigma N_p^{0}
	=  \mu_n,
	\qquad
	\beta N_p^{0}
	+
	\varsigma N_n^{0}
	=
	-\alpha.
\end{equation}
The mapping to~\eqref{eq_canonical_stationary_functional} is therefore given by the solution to the above system and the definitions
\begin{equation}
	\mathsf K 
	=
	\frac{\hbar^2}{2m^*}
	\begin{pmatrix}
		1 & 0 \\
		0 & 1 
	\end{pmatrix},
	\quad
	\mathsf\Lambda 
	=
	\begin{pmatrix}
		U_0       & \varsigma \\
		\varsigma & \beta     
	\end{pmatrix},
	\quad
	\mathsf C 
	= 0 \, .
\end{equation}
\\
\\
\emph{Relation with \citep{Shukla_PRD_2024}} -- 
\citet{Shukla_PRD_2024} consider a dynamical model for neutron and proton condensates coupled to the electromagnetic field and Newtonian gravity. To compare with the local stationary functional~\eqref{eq_canonical_stationary_functional}, we consider their non-rotating bulk sector.

After converting their electromagnetic convention to the one used here, their stationary free energy density reads
\begin{align}
	\mathcal F 
	={} &   
	\frac{\hbar^2}{2m_n}
	|\nabla\psi_n|^2
	+
	\frac{\hbar^2}{2m_p}
	|D_p\psi_p|^2
	+
	\frac{|\nabla\times\bm A|^2}{8\pi}
	\notag\\
	    & + 
	\frac{g }{2}
	\left(
	|\psi_n|^2-\frac{\mu_n}{g }
	\right)^2
	+
	\frac{\alpha_s}{2}
	\left(
	|\psi_p|^2-\frac{\mu_p}{\alpha_s}
	\right)^2
	\notag\\
	    & + 
	\gamma g_{np}|\psi_n|^2|\psi_p|^2
	-
	\gamma\,\bm J_n\cdot\bm J_p ,
	\label{eq_shukla_stationary}
\end{align}
where 
\begin{equation}
	\begin{split}
		& \bm J_n
		=
		\frac{\hbar}{2i}
		\left(
		\psi_n^*\nabla\psi_n
		-
		\psi_n\nabla\psi_n^*
		\right),
		\\
		& \bm J_p
		=
		\frac{\hbar}{2i}
		\left(
		\psi_p^*\nabla\psi_p
		-
		\psi_p\nabla\psi_p^*
		\right)
		-
		\frac{q}{c}\bm A|\psi_p|^2 .
	\end{split}
\end{equation}
The phase-gradient entrainment term is thus removed by setting $\gamma=0$, so that the model can be mapped into~\eqref{eq_canonical_stationary_functional}. However, this would also remove the $g_{np}$ terms, so we define
\begin{equation}
	\varsigma 
	=
	\gamma g_{np}.
\end{equation}
and take the double limit $\gamma\rightarrow0$,  $g_{np}\rightarrow\infty$ while keeping $\varsigma $ finite: $\varsigma < 0$ corresponds to the locally attractive case discussed in Sec.~\ref{sec_rep_attr}.
Imposing $\Psi_n=\psi_n$ and $\Psi_p=\psi_p$, the homogeneous equilibrium amplitudes satisfy
\begin{equation}
	g N_n^{0} +\varsigma N_p^{0} = \mu_n,
	\qquad
	\alpha_sN_p^{0}+\varsigma N_n^{0} = \mu_p.
\end{equation}
Therefore, the mapping to~\eqref{eq_canonical_stationary_functional} is realised by the solution to the above system and
\begin{equation}
	\mathsf K
	=
	\begin{pmatrix}
		\dfrac{\hbar^2}{2m_n} & 0                     
		\\[2mm]
		0                     & \dfrac{\hbar^2}{2m_p} 
	\end{pmatrix},
	\quad
	\mathsf\Lambda
	=
	\begin{pmatrix}
		g         & \varsigma 
		\\
		\varsigma & \alpha_s  
	\end{pmatrix},
	\quad
	\mathsf C=0 .
\end{equation}

\newpage

\bibliography{biblio}

\end{document}